\documentclass[
preprint,
superscriptaddress,
nofootinbib,
nobibnotes,
amsmath,amssymb,
prl,
]{revtex4-2}

\usepackage{graphicx}
\usepackage{dcolumn}
\usepackage{bm}
\usepackage{comment}
\usepackage[none]{hyphenat}
\usepackage{hyperref}
\hypersetup{colorlinks=true,citecolor=magenta,linkcolor=blue,breaklinks=true}
\usepackage{xcolor}
\usepackage{setspace}

\usepackage{wrapfig}
\usepackage{physics}

\newcommand{\mr}{\mathrm}

\begin{document}
\title{Superconducting magic-angle twisted trilayer graphene hosts \\competing magnetic order and moir\'e inhomogeneities}

\author{Ayshi Mukherjee}
 \homepage{Equal contribution}
 \email{ayshimukherjee@gmail.com}
 \author{Surat Layek}
 \homepage{Equal contribution}
 \affiliation{Department of Condensed Matter Physics and Materials Science, Tata Institute of Fundamental Research, Homi Bhabha Road, Mumbai 400005, India.}
\author{Subhajit Sinha}
 \email{sinhasubhajit25@gmail.com}
\affiliation{Department of Condensed Matter Physics and Materials Science, Tata Institute of Fundamental Research, Homi Bhabha Road, Mumbai 400005, India.}
\author{Ritajit Kundu}
\affiliation{Department of Physics, Indian Institute of Technology Kanpur, Kanpur 208016, India.}
\author{Alisha H. Marchawala}
\author{Mahesh Hingankar}
\author{Joydip Sarkar}
\author{L.D. Varma Sangani}
\author{Heena Agarwal}
\affiliation{Department of Condensed Matter Physics and Materials Science, Tata Institute of Fundamental Research, Homi Bhabha Road, Mumbai 400005, India.}
\author{Sanat Ghosh}
\author{Aya Batoul Tazi}
\affiliation{Department of Physics, Columbia University, New York, NY 10027, USA.}
\author{Kenji Watanabe}
\affiliation{Research Center for Functional Materials,
National Institute for Materials Science, 1-1 Namiki, Tsukuba 305-0044, Japan.}
\author{Takashi Taniguchi}
\affiliation{International Center for Materials Nanoarchitectonics, National Institute for Materials Science,  1-1 Namiki, Tsukuba 305-0044, Japan.}
\author{Abhay N. Pasupathy}
\affiliation{Department of Physics, Columbia University, New York, NY 10027, USA.}
\author{Arijit Kundu}
\affiliation{Department of Physics, Indian Institute of Technology Kanpur, Kanpur 208016, India.}
\author{Mandar M. Deshmukh}
\email{deshmukh@tifr.res.in}
\affiliation{Department of Condensed Matter Physics and Materials Science, Tata Institute of Fundamental Research, Homi Bhabha Road, Mumbai 400005, India.}

\begin{abstract}
{
The microscopic mechanism of superconductivity in the magic-angle twisted graphene family, including magic-angle twisted trilayer graphene (MATTG), is poorly understood. Properties of MATTG, like Pauli limit violation, suggest unconventional superconductivity. Theoretical studies propose proximal magnetic states in the phase diagram, but direct experimental evidence is lacking. We show direct evidence for an in-plane magnetic order proximal to the superconducting state using two complementary electrical transport measurements. First,  we probe the superconducting phase by using statistically significant switching events from superconducting to the dissipative state of MATTG. The system behaves like a network of Josephson junctions due to lattice relaxation-induced moir\'e inhomogeneity in the system. We observe non-monotonic and hysteretic responses in the switching distributions as a function of temperature and in-plane magnetic field. Second, in normal regions doped slightly away from the superconducting regime, we observe hysteresis in magnetoresistance with an in-plane magnetic field; showing evidence for in-plane magnetic order that vanishes $\sim$900 mK. Additionally, we show a broadened Berezinskii–Kosterlitz–Thouless transition due to relaxation-induced moir\'e inhomogeneity. We find superfluid stiffness $J_{\mathrm{s}}$$\sim$0.15~K with strong temperature dependence. Theoretically, the magnetic and superconducting order arising from the magnetic order’s fluctuations have been proposed – we show direct evidence for both. Our observation that the hysteretic magnetoresistance is sensitive to the in-plane field may constrain possible intervalley-coherent magnetic orders and the resulting superconductivity that arises from its fluctuations.

} 
\end{abstract}
\maketitle
 
The family of twisted multilayer graphene devices, like twisted bilayer and trilayer graphene, provide an opportunity to study the origin of superconductivity (SC) in these materials hosting flatbands \cite{park2021tunable,hao2021electric,park2022robust}. The magic-angle twisted trilayer graphene (MATTG) hosts both the Dirac band and moir\'{e} flatband \cite{phong2021band,li2022observation}, and exhibits Pauli limit violation \cite{park2022robust,lake2021reentrant}. Such exotic properties make MATTG an interesting system to study. Recent microscopic theoretical studies show that the superconducting regions are surrounded, in the phase diagram, by phases and ordering of different kinds \cite{fischer2022unconventional,christos2022correlated,gonzalez2023ising,zhang2024angle}, making spin (valley) configuration and spin (valley) fluctuations an important physics in the system. Co-existence of different phases may give rise to competition between them \cite{classen2019competing}.

The stacking of atomically thin sheets of graphene with a twist angle gives rise to lattice reconstruction and lattice relaxation effects. In general, moir\'e systems can also incur moir\'e of moir\'e superstructure domains due to strong lattice relaxation \cite{nakatsuji2023multiscale}. Studies using STM techniques have revealed the presence of quasi-one dimensional `moir\'e solitons' and point-like faults `twistons' in MATTG \cite{turkel2022orderly, kim2022evidence} due to moir\'e lattice reconstruction. The presence of such relaxation-induced moir\'e inhomogeneities distinct from twist angle disorder can make the understanding of underlying mechanisms more challenging. Intrinsic mesoscopic inhomogeneities also exist in systems like LAO/STO \cite{prawiroatmodjo2016evidence,hurand2019josephson} exhibiting broadened Berezinskii– Kosterlitz–Thouless (BKT) transition \cite{benfatto2009broadening,venditti2019nonlinear}. The measurement of superfluid stiffness using BKT-like analysis is a route to extract important information about the superconductivity in MATTG and moir\'e inhomogeneity makes it challenging.

In this article, we report evidence of competing order and moir\'e inhomogeneity in the superconducting phase of MATTG, through quantum transport and switching measurements in the superconducting and neighboring normal states. We present the switching measurements as a new approach to understanding these superconductors. They are studied with both temperature and in-plane magnetic fields to understand the system's spin configuration and ground state. We report a non-monotonic behavior of the switching distributions with temperature strongly pointing towards a competing order, likely  magnetic in origin, in the ground state. The switching measurement is largely successful in bringing out exciting features in the system as the mesoscopic moir\'e inhomogeneity in the system allows us to describe it as an array of Josephson junctions (JJs). The switching distribution points towards evidence of a magnetic order when probed with an in-plane magnetic field. The observation of hysteresis, in the proximal normal phase, in magnetoresistance with in-plane magnetic field provides direct evidence to support the results of switching measurements. Our experiments and analysis also provide a way to infer the difficult-to-measure quantity of superfluid stiffness - a first estimate in the system of MATTG; which reveals that the observation of BKT transition in the system is broadened due to moir\'e inhomogeneity.

The MATTG is a mirror symmetric stacking of three layers of graphene ~- with the middle layer twisted by the magic angle. (See Supplementary Information sec. II for different twist angles realized in devices.) The stack is encapsulated in hBN and has a top gate and bottom gate. The dual-gate geometry allows independent control over charge density $n$ and applied perpendicular electric displacement field $D$. (See Supplementary Information sec. III and IV for fabrication and measurement details, respectively.)

\begin{figure*}[h!]
    \centering
    \includegraphics{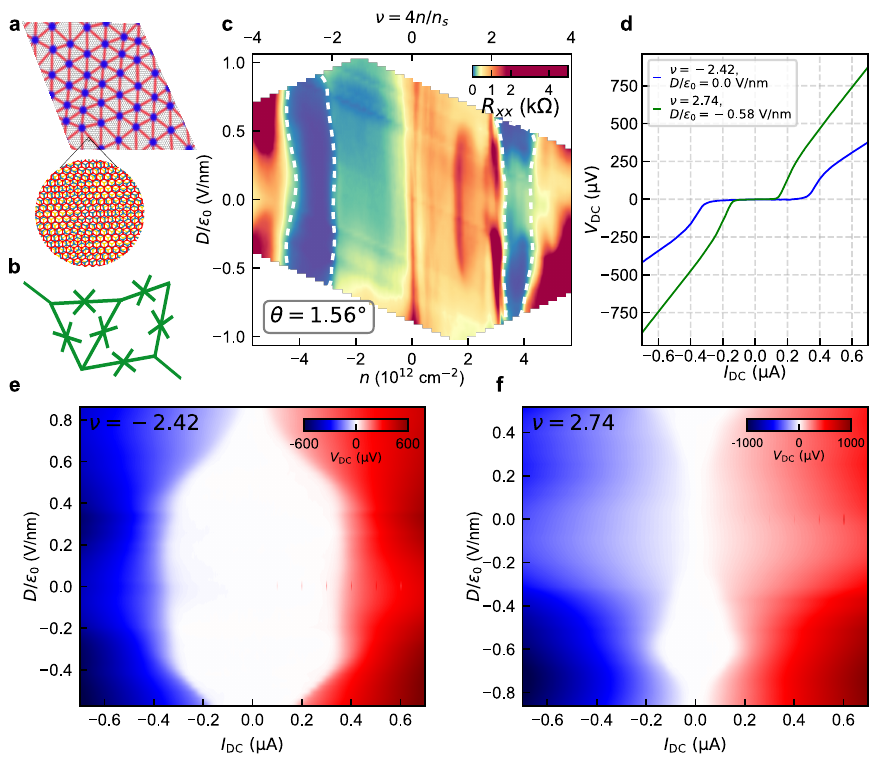}
    \caption{ \label{fig:fig1} { \textbf{Superconductivity in MATTG.}
    \textbf{a,}~ Schematic depiction of moiré lattice reconstruction, leading to the formation of an array of twistons (shaded in blue) and moiré solitons (shaded in red), featuring local twist-angle faults. Inset - a close-up perspective of the MATTG moiré, illustrating the contrasting length scales of solitons and moiré.
    \textbf{b,}~In the superconducting state, moiré twistons and solitons in a twisted trilayer graphene serve as weak links within the superconductor, forming a network of Josephson junctions.
    \textbf{c,}~ Longitudinal resistance $R\mathrm{_{xx}}$ as a function of carrier density $n$, filling $\nu$, and electric field $D/\epsilon_0$ at $T$=20~mK, $B=0$~T. The superconducting regions have been marked by a white dashed outline.
    \textbf{d,}~DC voltage drop across the device as a function of DC current bias at the optimal hole and electron doping. The hole-doped side and electron-doped side have a maximum critical current of $\sim$400~nA and $\sim$200~nA, respectively. 
    \textbf{e, f,}~DC $V\mathrm{_{DC}}$-$I\mathrm{_{\mathrm{DC}}}$ curves at optimal hole (e) and electron (f) doping with varying $D/\epsilon_0$. The white part represents zero voltage drop – superconducting part before it turns normal (red/blue). The boundary of the white region gives us an idea about the critical current, at a particular $D/\epsilon_0$.
    }}
\end{figure*}

MATTG has been reported as a robust superconductor having a $T_\mathrm{c}\sim 2$~K with predictions of exotic superconducting properties like spin-triplet superconductivity \cite{cao2021pauli,lake2021reentrant,christos2022correlated,cao2024robust,choi2021dichotomy,chou2021correlation}. Fig.~\ref{fig:fig1}a schematically shows the formation and manifestation of twistons and moir\'e solitons in the system of MATTG. The plaquette regions have a twist angle that is close to the magic angle of ~1.56$^\circ$, whereas the twiston and soliton regions have higher twist angles due to lattice relaxation \cite{turkel2022orderly}. Such twist angle variation among the regions lead to different moiré lengthscales and, in turn, gives rise to variations in the local filling factor. The twiston and soliton regions have smaller filling factors and thus act as weak links to the superconducting plaquettes. (Details of variation of the density of states with the local twist angle from non-interacting theory is in the Supplementary Information sec. I.) Weak links distributed in the system lead to the formation of a network of JJs as shown in Fig.~\ref{fig:fig1}b which is discussed in detail later. We report that MATTG hosts SC in both the hole-doped regime and electron-doped regime -- consistent with past studies. The filling~$\nu~(=4n/n_{s}$, where $n_{\mathrm{s}}=5.67\times10^{12}~\mathrm{cm^{-2}}$ is the superlattice density) and $D/\epsilon_0$ where the SC emerges is seen in Fig.~\ref{fig:fig1}c, representing zero longitudinal resistance R$\mathrm{_{xx}}$ around filling $\nu=\pm 3$. We note critical temperatures around 1.6 K and 1.2 K in the hole-doped regime and electron-doped regime, respectively. (See Supplementary Information sec. V.) Fig.~\ref{fig:fig1}d shows DC $V\mathrm{_{DC}}$-$I\mathrm{_{DC}}$ curves at optimal hole and electron fillings. We find a critical current of around 400~nA for optimal hole biasing of $\nu=-2.42,~D/\epsilon_0=0.0~\mathrm{V/nm}$, and 200~nA for optimal electron biasing of $\nu=2.74,~D/\epsilon_0=-0.58~\mathrm{V/nm}$ ~– comparable to other reported values in the literature \cite{park2021tunable,cao2021pauli,hao2021electric}. We also measure another device with a critical current of around 50~nA for optimal hole biasing (see Supplementary Information sec. VI).

The electric field $D/\epsilon_0$, can be tuned to study different phases in MATTG. Here, we study the modulation of the strength of SC in the system with $D/\epsilon_0$. Figs.~\ref{fig:fig1}e and f show the $V\mathrm{_{DC}}$-$I\mathrm{_{DC}}$ curves as a function of $D/\epsilon_0$ for optimal hole and electron-filling, respectively. The hole side superconducting phase is the strongest at zero electric field and weakens after an electric field of $\pm$ 0.50~V/nm. In contrast, the electron side superconducting phase hosts the maximum critical current at a finite electric field of 0.59~V/nm and is considerably weaker at 0~V/nm. (See Supplementary Information sec. VII for further device characterization.) Our observation of distinct dependence of superconductivity on the electric field in electron and hole fillings is consistent with past experiments \cite{fischer2022unconventional,christos2022correlated,gonzalez2023ising}. We next discuss a method of studying SC that has not been used for twistronic superconductors. 

\begin{figure*}[h!]
    \centering
    \includegraphics{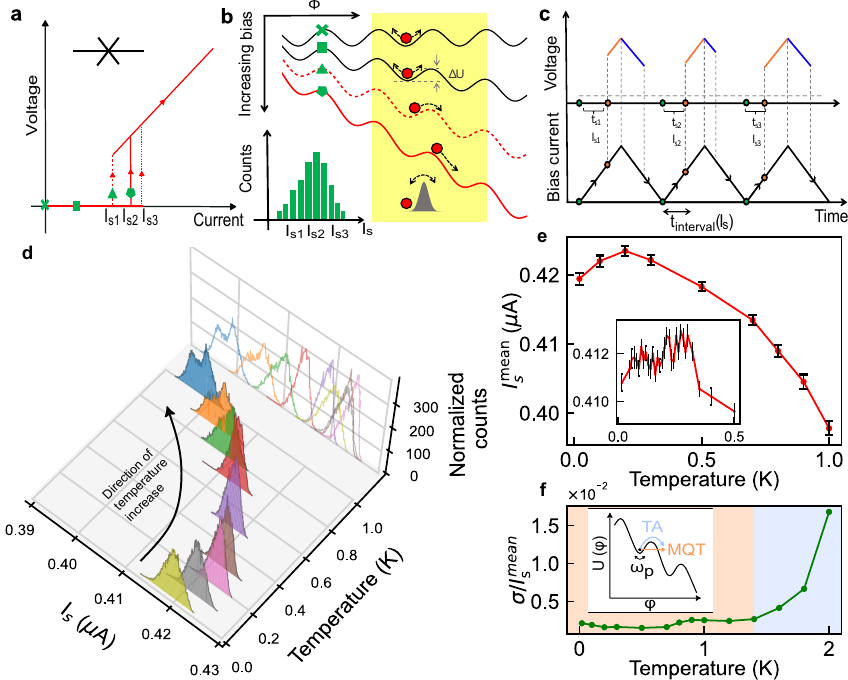}
    \caption{
        \label{fig:fig2}
        \textbf{Switching statistics with temperature suggests inhomogeneities and competing order.}
            {\scriptsize
            \textbf{a,}~ The DC I-V characteristics of a Josephson junction (JJ) showing the different switching currents. 
            \textbf{b,}~The tilted washboard potential captures the JJ's switching from superconducting to normal state by an analogous picture of a particle crossing the potential barrier $\Delta U$. Larger current biases tilt the washboard potential more, making it easier for the particle to cross the now-tilted barrier. Inset - the switching histogram showcases the stochastic nature of the JJ switching at different switching currents.
            \textbf{c,}~ To gather statistics of the switching current, a low-frequency triangular wave is applied by a function generator. The current undergoes a linear variation with time, starting from zero and reaching a value slightly above the critical current. The counter measures the elapsed time ($t_{\mathrm{s}}$) from the onset of zero bias current (green dot) to the instance of transition from the superconducting to the normal state (red dot). These switching times are then utilized to determine the switching currents ($I_{\mathrm{s}}$), taking into account the frequency and amplitude of the triangular signal. 
            \textbf{d,}~Switching histograms at optimal hole-side doping and electric field  ($\nu$, $D/\epsilon_0$) = (-2.42, 0) as a function of temperature. The distributions evolve non-monotonically with temperature. The arrow represents the direction of the temperature sweep.
            \textbf{e,}~The mean current of the switching distributions $I_{s}^{\mathrm{mean}}$ plotted as a function of temperature. The $I_{s}^{\mathrm{mean}}$ increases from 20~mK to 200~mK and decreases thereafter, which is a non-monotonic behavior. Inset - $I_{s}^{\mathrm{mean}}$ plotted with temperature bin of 10 mK clearly shows the non-monotonic behavior at ($\nu$, $D/\epsilon_0$) = (-2.42, -0.29). The error bars denote the standard deviations of individual distributions.
            \textbf{f,}~The standard deviation normalized by the mean of the distributions $\sigma/I_{\mathrm{s}}^{\mathrm{mean}}$, plotted as a function of temperature. The trend of evolution of $\sigma/I_{\mathrm{s}}^{\mathrm{mean}}$ with temperature denotes the different switching processes dominating at different temperatures. The initial temperature-independent macroscopic quantum tunneling (MQT) process (orange) transitions into the thermal activation (TA) process (blue) where the $\sigma/I_{\mathrm{s}}^{\mathrm{mean}}$ increases. Inset - The washboard potential U varies as a function of phase $\Phi$. The
            different switching processes – TA and MQT are shown schematically where $\omega_{p}$ is the frequency of oscillation of the particle in the potential well.
            }
    }
\end{figure*}

The switching measurements capturing the transition from a superconducting to a dissipative state bring out the stochastic nature of the switching current $I_{\mathrm{s}}$ ~– one that is not apparent in a single $V\mathrm{_{DC}}$-$I_{\mathrm{DC}}$ measurement. The switching of JJs being a stochastic process leads to the switching taking place at different bias currents as schematically depicted in Fig.~\ref{fig:fig2}a. Fig.~\ref{fig:fig2}b shows a schematic of the washboard potential landscape associated with the resistively capacitance shunted junction (RCSJ) model which captures the dynamics of a JJ. On increasing the current bias, the potential tilts, and the particle in the landscape can escape the potential barrier. The escape is analogous to the JJ switching from the superconducting state to the normal state.  The stochastic nature of the switching of JJs is captured in the switching current histogram as shown in the inset of Fig.~\ref{fig:fig2}b.  Each histogram showcases large statistics that help provide insight into the nature of the superconducting transition, inhomogeneities in the system, and energetics \cite{wallraff2003switching,sahu2009individual}. We record 10,000 switching events to gather a normalized histogram distribution of $I_{s}$. Fig.~\ref{fig:fig2}c presents the principle of the switching current measurement technique which allows us to gather large statistics about the stochastic quantity of switching current $I_{\mathrm{s}}$. (See Supplementary Information sec. VIII for details of the switching measurements.)

Fig.~\ref{fig:fig2}d shows the switching distributions, as a function of temperature, at the optimal hole filling. 
The mean switching current $I_{s}^{\mathrm{mean}}$ of the distribution has a non-monotonic behavior~ – increases with temperature up to $\sim$200~mK and thereafter decreases in Fig.~\ref{fig:fig2}e. This non-monotonic response is not expected for conventional JJs. A possible way for the increase in $I_{s}^{\mathrm{mean}}$ with temperature up to 200~mK is the suppression of a competing order to give way to the superconducting state – noted as an enhancement of the critical current \cite{fischer2022unconventional,christos2022correlated}. (See Supplementary Information sec. IX for a simple phenomenological model.) A non-monotonic evolution of $I_{s}^{\mathrm{mean}}$ is also noted in the electron-doped SC. (See Supplementary Information sec. X for additional data.) Such an enhancement in critical current has been reported previously for magnetic-ordered materials \cite{weigand2013strong}, and materials having d-wave superconducting order parameter \cite{iguchi1994experimental}. This forms our motivation to look for a competing order that is magnetic in nature. Here, we provide the first indication of the competing magnetic ground states in MATTG where the superconducting order couples with a magnetic order \cite{ramires2021emulating} within an energy scale of $\sim$200~mK. Aspects of this competing order will be further seen in measurements with in-plane magnetic fields later in this article. First, we note the evolution of switching distribution beyond 200~mK in Fig.~\ref{fig:fig2}d when the distributions get wider and subsequently develop substructure -- as we discuss next, this provides insight into the spatial structure of the superconductor. 

Spatial inhomogeneities in  LAO/STO system lead to the creation of weak links between the superconducting parts of the system – which in turn create an array of Josephson junctions (JJs) \cite{hurand2019josephson}. This description of an array of JJs is also suitable to MATTG owing to the moir\'e inhomogeneities present in the system \cite{turkel2022orderly} and is further supported by our switching measurements that illustrate the stochastic nature of $I_{s}$ which is a characteristic of JJs \cite{fulton1974lifetime,van1991dynamics,van1993vortex}. The array of the JJ-coupled superconducting islands can be modeled to an equivalent resistively capacitance shunted junction (RCSJ) circuit. The particle in the washboard potential landscape can escape either by a macroscopic quantum tunneling (MQT) or a thermal activation (TA) process (see Fig. \ref{fig:fig2}f inset). The system can undergo a transition to a lower energy state by thermal excitations over the intervening barrier at sufficiently high temperatures. However, at lower temperatures transition across the barrier occurs via quantum mechanical tunneling - a process independent of temperature. The histograms' standard deviation ($\sigma$) showcases this temperature dependence for both processes, allowing us to extract the microscopic information. The categorization of the system in either of these regimes is done by noting the standard deviation divided by the mean switching current $\sigma/I_{\mathrm{s}}^{\mathrm{mean}}$ of the switching distribution as a function of temperature – plotted in Fig.~\ref{fig:fig2}f. We can note that $\sigma/I_{\mathrm{s}}^{\mathrm{mean}}$ remains weakly dependent on temperature, a characteristic of MQT, up to 1 K. Thereafter, $\sigma/I_{\mathrm{s}}^{\mathrm{mean}}$ increases with temperature indicating a transition into the TA regime. The temperature of this transition is called the cross-over temperature $T_{\mathrm{CO}}$ and is around 1 K for this system. (See Supplementary Information sec. XII for additional measurements showing the suppression of $T_{\mathrm{CO}}$ on the application of a small finite perpendicular magnetic field.) A similar transition from MQT to TA regime is also noted in graphene-based JJs \cite{lee2011electrically}. We also use the RCSJ model to analyze the MATTG as an array of JJs, and estimate the shunt capacitance $C$ to be $C\simeq 1.3$~fF (Details in Supplementary Information sec. XI). This value of capacitance allows an independent cross-check of the $T_{\mathrm{CO}}$ in the system. (See Supplementary Information sec. XII.)  
All these evidences show that we can visualize the system as a JJ network.

The switching distribution histograms also develop a substructure at temperatures above 1 K – peaked at two current values (blue colored histogram in Fig.~\ref{fig:fig2}d). The different twist angles in the device host different $I_{s}^{\mathrm{mean}}$ due to relaxation-induced moir\'e inhomogeneities and are separated out by increasing the temperature resulting in the double-peaked distribution. As the temperature increases further, beyond 1 K, the bimodal distribution gradually evolves into a broad distribution without much substructure at 2 K. (See Supplementary Information sec. X D.) The bimodal distributions seen at higher temperatures further add to the list of evidence in support of the claim that MATTG is an inhomogeneous superconductor with regions of superconductor separated by moiré solitons and twistons \cite{turkel2022orderly}. This inhomogeneous nature of the system, we believe, could only be brought out by studying large statistics of switching events.

\begin{figure*}[h!]
    \centering
    \includegraphics[width=17.2cm]{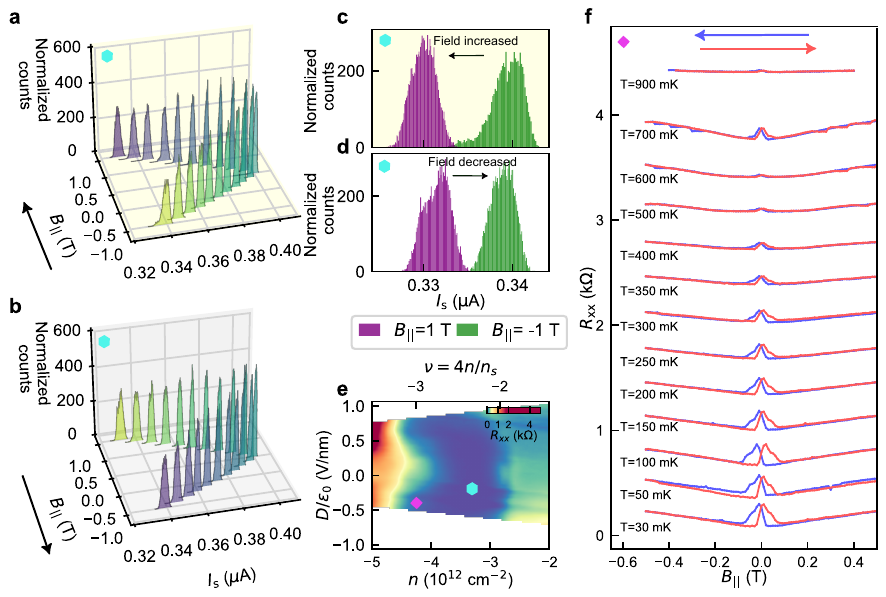}
    \caption{ 
    \label{fig:fig3} 
    \textbf{Switching statistics and hysteresis with magnetic field suggest competing magnetic order.}
    {
    \textbf{a, b,}~Switching histograms at optimal hole-side doping and electric field at $T$=20 mK, as a function of the in-plane magnetic field $B_{\parallel}$ up to $\pm$~1~T, in (a) increasing field sweep direction and (b) decreasing field sweep direction. The distributions show magnetic field direction-dependent response. The arrows represent the direction of $B_{\parallel}$ sweep.
    \textbf{c, d,}~The switching histograms at 1 T and  - 1 T for the (c) increasing field sweep and (d) decreasing field sweep pointing out the direction-dependent response of the device to $B_{\mathrm{||}}$ in that the switching current $I_{\mathrm{s}}$ is different for 1 T and -1 T. 
    \textbf{e,}~Plot of the longitudinal resistance $R_{\mathrm{xx}}$ in the $n$-$D$ parameter space, zoomed in near the hole-side superconducting region. (a-d) switching data is taken at the hole-side superconducting region marked by cyan hexagon. Whereas, a normal ($R_{\mathrm{xx}}\neq 0$) region is marked by a magenta diamond, and identifies the doping and electric field at which the data in (f) is acquired.
    \textbf{f,}~Longitudinal resistance $R_{\mathrm{xx}}$ plotted as a function of in-plane magnetic field $B_{\mathrm{||}}$, at a doping and electric field identified by a magenta diamond in (e) and marks the phase in the neighborhood of the superconducting phase. We observe butterfly hysteresis features that evolve and subsequently vanish with increasing temperature. The arrows specify the direction of $B_{\mathrm{||}}$ sweep. Plots at each temperature are shifted along the y-axis for clarity.  }
    }
\end{figure*}

As we discussed earlier, we observe a non-monotonic variation of $I_{s}^{\mathrm{mean}}$ with temperature - an aspect we attributed to possible competition between the superconducting and magnetic order. We now probe the possibility of magnetic order using an in-plane magnetic field $B_{\parallel}$ \cite{qin2021plane} by studying the evolution of the switching distribution in $B_{\parallel}$. Fig.~\ref{fig:fig3}a and b shows the switching histograms from 1~T to -1~T and vice versa, plotted at a bin of 0.1~T. The distributions for 1~T and -1~T magnetic fields differ from each other in their $I_{s}^{\mathrm{mean}}$ for both directions of field sweep as shown in Fig.~\ref{fig:fig3}c and d. This brings out the striking difference in the response of the system to the direction of the magnetic field -- second evidence for competing order that we provide using switching measurements. (See Supplementary Information sec. XIII for additional thermal cycling data of switching histograms in presence and absence of in-plane magnetic field.) Such a behavior can be attributed to a combination of spin-singlet and triplet configuration in the system \cite{lake2021reentrant}, or competing orders in the vicinity of the superconducting order \cite{fischer2022unconventional,ramires2021emulating,christos2022correlated,classen2019competing}; making MATTG a potential platform to study competition between a superconducting and magnetic order. 

Fig.~\ref{fig:fig3}e marks the superconducting region where the switching experiments are performed and the neighboring region where a magnetic order is likely present -- this uses our second technique distinct from switching measurements.  It is interesting to note from Fig.~\ref{fig:fig3}f that the longitudinal resistance $R_{\mathrm{xx}}$ in the vicinity of the superconducting phase in the $n$-$D$ phase diagram shows a hysteretic behavior with in-plane magnetic field $B_{\mathrm{||}}$. This shows strong evidence of a magnetic order that is present in the vicinity of the superconducting state that couples preferentially to an in-plane magnetic field. The hysteretic response dies out with temperature and is absent from 900 mK. We do not however understand the re-emergence of hysteresis at 700 mK. 
Next, we discuss the key aspects of our observation of magnetic hysteresis in longitudinal resistance. 

In Fig.~\ref{fig:fig3}f, we see the hysteresis in longitudinal resistance as one dopes the system away from the superconducting region (see Extended Data Fig.~1). Firstly, the hysteresis is pronounced with an in-plane magnetic field while it is subtle but observable with a perpendicular magnetic field (see Extended Data Fig.~1 for $R_{\mathrm{xx}}$, and Supplementary Information Sec.~XIV for Hall resistance). Secondly, the hysteresis with an in-plane magnetic field is accompanied by substantial magnetoresistance ($\sim 50$\% for a field up to $\sim$ 0.5~T). Lastly, the hysteresis and the magnetoresistance disappear at temperatures $\sim$~900~mK (see Extended Data Fig.~4), indicating a common origin for distinct observations of hysteresis and magnetoresistance.  Additionally, in samples that do not show the superconducting response, the hysteresis, and magnetoresistance are still present (see Extended Data Fig.~2 and 3). Our switching data, together with the longitudinal magnetoresistance provides strong evidence for the proximal magnetic and superconducting order.

Additionally, we observe the superconducting diode effect in our MATTG devices with a current asymmetry of $\sim$1.2\% (see Supplementary Information sec. XV) suggesting time-reversal symmetry breaking consistent with our direct observations. The non-zero asymmetry is consistent with the presence of competing magnetic order as we discuss next \cite{banerjee2024enhanced}. The superconducting diode effect refers to the asymmetry in $I_{\mathrm{s}}$ for the positive and negative current biases; it suggests the breaking of time-reversal and inversion symmetries.  Lin et al. report zero-field diode effect in twisted trilayer graphene on WSe$_{2}$ heterostructure  \cite{lin2022zero}. Our observation is consistent with past works that indicate time-reversal symmetry breaking in MATTG, which we claim is due to the presence of magnetic order.

\begin{figure*}[h]
    \centering
    \includegraphics{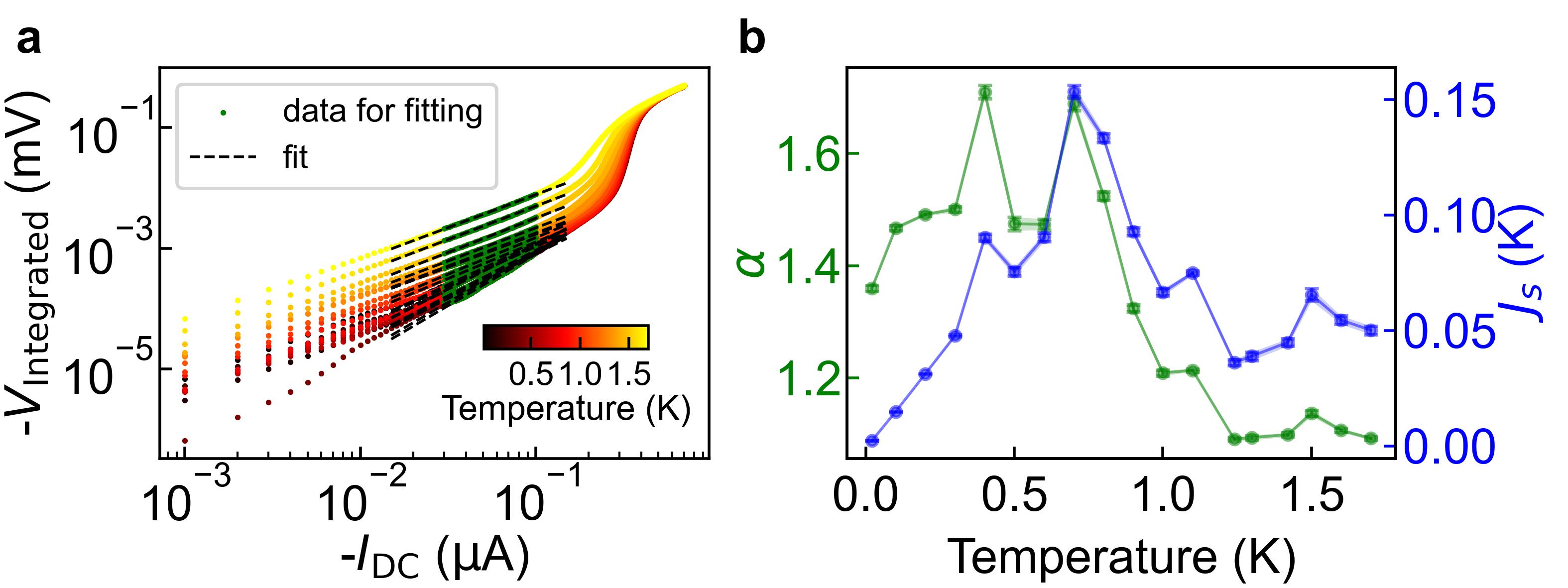}
    \caption{ \label{fig:fig4} {\textbf{Superfluid stiffness estimation showing a broadened BKT transition.}
        \textbf{a,}~Integrated $dV/dI$ curves to produce $V_{\mathrm{Integrated}}-I_{\mathrm{DC}}$ curves, at different temperatures; are separate measurements from the switching measurements. Fits are done for the non-linear exponent at low current values of around 50 nA.
    \textbf{b,}~Extracted exponent $\alpha$ values and $J_{\mathrm{s}}$ as a function of temperature. The exponents are extracted from the fits in (a). The $\alpha$ values do not show a sharp transition. The fitting error bars are comparable to individual data point markers.
    }
    }
\end{figure*}
MATTG is a 2D system and is expected to undergo a BKT transition, derived within the clean XY model; this forms the motivation to study the evolution of the superconducting phase in MATTG. We expect to observe a discontinuous jump in superfluid stiffness $J_{\mathrm{s}}(T_{\mathrm{BKT}}^{-})=\frac{2}{\pi}T_{\mathrm{BKT}}, ~ J_{\mathrm{s}}(T_{\mathrm{BKT}}^{+})=0$ in the clean limit, where $T_{\mathrm{BKT}}$ is the BKT transition temperature and $T_{\mathrm{BKT}}^{-}$ and $T_{\mathrm{BKT}}^{+}$ are temperatures just before and after the transition. It is also related to a non-linear exponent in the $V\mathrm{_{\mathrm{DC}}}$-$I\mathrm{_{DC}}$ curves and allows us to estimate $J\mathrm{_{s}}$ in such systems \cite{venditti2019nonlinear}, 

\begin{eqnarray}
    V\propto I^{\alpha(T)},~ 
    \alpha(T)=1+\frac{\pi J_{s}(T)}{T}.
\end{eqnarray}

The corresponding jump in $\alpha$, $\alpha(T_{\mathrm{BKT}}^{-})=3,~ \alpha(T_{\mathrm{BKT}}^{+})=1$ is used to characterize the BKT transition temperature in 2D twisted graphene heterostructures exhibiting SC. However, this description holds true for a clean limit of the sample. In disordered samples, the $J\mathrm{_{s}}$ is strongly suppressed – giving rise to a fragile superconducting condensate, like has been reported in systems like LAO/STO. The disorder gives rise to spatially isolated puddles of superconducting regions and connects to the inhomogeneous superconductor picture, arising from relaxation-induced moir\'e inhomogeneities in MATTG presented earlier.

Estimating $J_{\mathrm{s}}$ in such systems with mesoscopic scale moir\'e inhomogeneities is challenging owing to the small dimension of samples as well as the broadening of the BKT transition due to the inhomogeneities \cite{benfatto2009broadening,venditti2019nonlinear}. An important and overlooked consideration in this analysis is that the current biases at which $\alpha$ is extracted must be about an order of magnitude smaller than the typical $I_{\mathrm{s}}$. At lower currents, the exponent captures the vortex-antivortex de-pairing central to BKT physics rather than the depairing of Cooper pairs close to $I_{\mathrm{s}}$. We extract $\alpha$ from $V_{\mathrm{integrated}}$ vs $I_{\mathrm{\mathrm{DC}}}$ curves at low currents (see Fig.~\ref{fig:fig4}a). (See Supplementary Information sec. XVI for details.) We observe a strong suppression of $\alpha$ (always less than 3) and no sharp transition in Fig.~\ref{fig:fig4}b. The values of $J\mathrm{_{s}}$ obtained are comparable to values reported for twisted bilayer graphene in Ref. \cite{tian2023evidence}. We do not fully understand the behavior of $J\mathrm{_{s}}$ with temperature however, the absence of a sharp transition suggests that moir\'e inhomogeneities indeed play an important role in the kind of physics observed in MATTG superconductivity.
 
 We use switching measurements to characterize the MATTG superconductor and the magnetoresistance in the proximal normal phase. Our technique gives direct insight into the spatial moir\'e inhomogeneities and the competing orders in the system \cite{fischer2022unconventional,christos2022correlated, ramires2021emulating,yu2023magic,batlle2024cryo}, reflected in the switching response and magnetoresistance as a function of temperature and in-plane magnetic field. While we show direct evidence of a magnetic competing order, the origin is clearly from an order that couples to the parallel magnetic field. The normal state hysteretic magnetic response could arise from localized moments. Our experiments provide credence for the heavy fermion description for the MATTG with the localized moment. In addition, our experimental findings will constrain the possible correlated magnetic states that emerge from the intervalley coherent order \cite{christos2022correlated}.
 As we advance, it may be possible to probe the quantum phase transition as one transits from the normal magnetic state to the superconducting state with coexisting magnetic order.

\subsection{Data Availability}
The data that support the current study are available from the corresponding authors upon reasonable request.

\subsection{Acknowledgements}

We thank José Lado, Lara Benfatto, Allan MacDonald, Sophie Guéron, Hélène Bouchiat, Rhine Samajdar, Felix von Oppen, Vladimir Krasnov, Shubhayu Chatterjee, and Siddharth A. Parameswaran for helpful discussions. We thank Soumyajit Samal, Rishiraj Rajkhowa, and Abhishek Sunamudi for assistance in fabrication. We thank Pratap Chandra Adak for providing inputs on fabrication. M.M.D. acknowledges Nanomission grant SR/NM/NS45/2016 and DST SUPRA SPR/2019/001247 grant along with the Department of Atomic Energy of Government of India 12-R\&D-TFR-5.10-0100 for support. M.M.D. acknowledges support from J.C. Bose Fellowship JCB/2022/000045 from the Department of Science and Technology of India. K.W. and T.T. acknowledge support from the Elemental Strategy Initiative conducted by the MEXT, Japan (grant no. JPMXP0112101001), and JSPS KAKENHI (grant nos. 19H05790 and JP20H00354).
R.K. acknowledges support from the PMRF fellowship.
A.K. acknowledges support
from the SERB (Govt. of India) via sanction No.
CRG/2020/00180. 

\subsection{Author contributions}

A.M. fabricated the samples. A.M. and S.L. performed the measurements and analyzed the data. S.S. helped with measurements and analysis. R.K. and A.K. did the theoretical calculations. S.L., A.H.M., M.H., H.A., L.D.V.S., S.G., and A.B.T. helped in fabrication. J.S. helped with measurements. K.W. and T.T. grew the hBN crystals. A.N.P. gave inputs on fabrication. A.M. and M.M.D. wrote the manuscript with input from all authors. M.M.D. supervised the project.

\subsection{Competing Interests}
The authors declare no competing interests.

\bibliography{apssamp}
\newpage
\appendix
\setcounter{figure}{0}
\renewcommand{\figurename}{Extended Data Fig.}
\renewcommand{\thefigure}{\arabic{figure}}

\begin{figure*}[h!]
    \centering
    \includegraphics{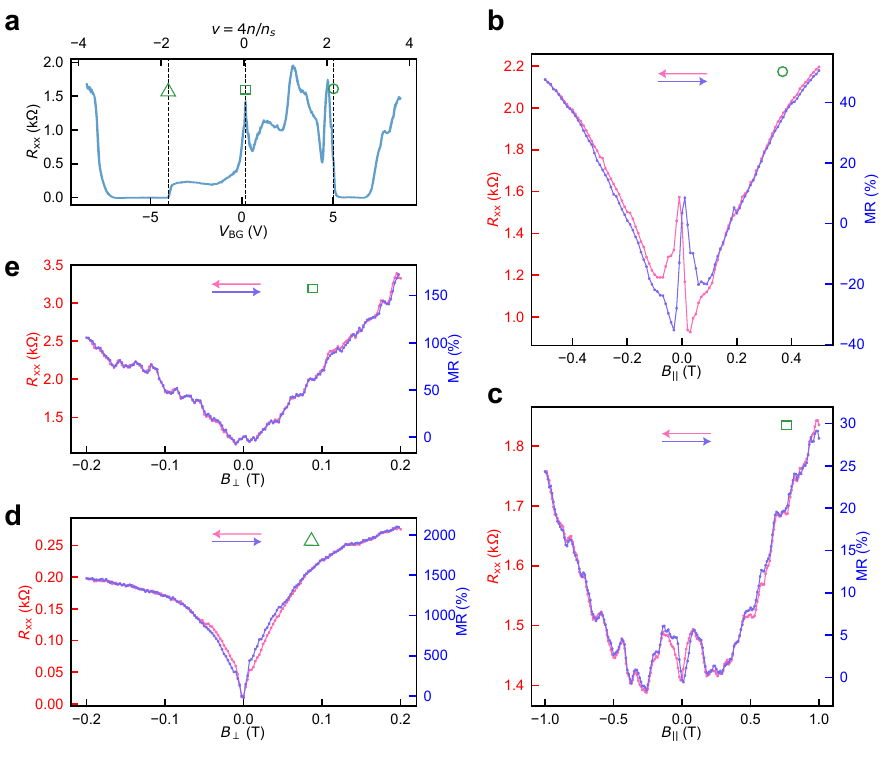}
    \caption{ \label{fig:fig6} 
    {\textbf{Magnetoresistance and hysteresis at additional filling factors and with in-plane and perpendicular magnetic field.}
    \textbf{a,} Lineslice of longitudinal resistance $R_{\mathrm{xx}}$ as a function of bottom gate voltage $V_{\mathrm{BG}}$ (the top axis indicates the filling factor $\nu$) in device A. The dashed lines identify the dopings at which the following magnetoresistance measurements are done. The triangle, square, and circle represent a hole-doped bias, a charge neutrality point (CNP) bias, and an electron-doped bias respectively.
    \textbf{b, c,} The longitudinal resistance $R_{\mathrm{xx}}$ and magnetoresistance $(\mathrm{MR}= [R_{\mathrm{xx}}(B)-R_{\mathrm{xx}}(B=0)]/R_{\mathrm{xx}}(B=0))$ as a function of in-plane magnetic field $B_{\mathrm{||}}$ at dopings identified by the symbols on the upper right-hand corner. The hysteresis is relatively less prominent at the CNP.
    \textbf{d, e,} $R_{\mathrm{xx}}$ and MR as a function of perpendicular magnetic field $B_{\mathrm{\perp}}$ at dopings identified by the symbols on the upper right-hand corner. The hysteresis with $B_{\mathrm{\perp}}$ is not as striking as with $B_{\mathrm{||}}$ however, it is present in  subpanel \textbf{d}.
    The arrows indicate the direction of the magnetic field sweep. This serves as a direct proof that indeed the superconducting phase is surrounded by phases that have magnetic
ordering. We further present Hall resistance hysteresis data in the Supplementary Information sec. XIV as additional evidence of broken time-reversal symmetry due to proximal magnetic order. 
    }
    }
\end{figure*}
\newpage

\begin{figure*}[h!]
    \centering
    \includegraphics{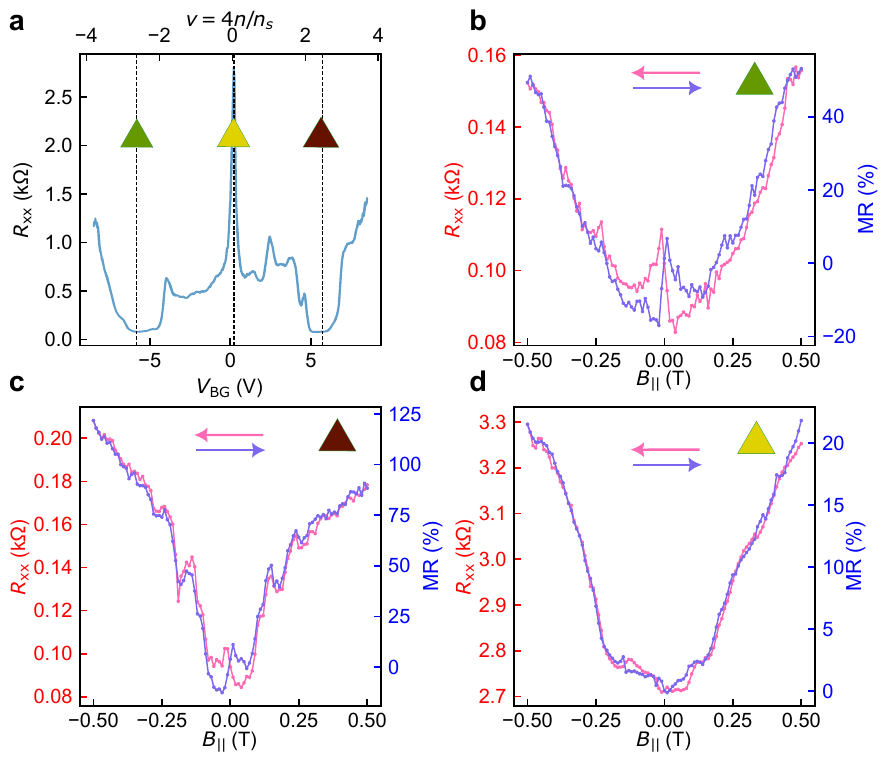}
    \caption{ \label{fig:fig7} 
    {\textbf{Magnetoresistance and hysteresis in a spatially adjacent non-superconducting region with angle $1.44^{\circ}$ at T= 20 mK. }
    \textbf{a,} Lineslice of longitudinal resistance $R_{\mathrm{xx}}$ as a function of bottom gate voltage $V_{\mathrm{BG}}$ (the top axis indicates the filling factor $\nu$) in device A for a spatially adjacent probe that does not show superconductivity but shows the characteristic dip in $R_{\mathrm{xx}}$, which drops to a small non-zero value. The motivation is to look for hysteresis at the same dopings that show $R_{\mathrm{xx}}=0$ for the spatially adjacent superconducting probes in the same device. The dashed lines identify the dopings at which the following magnetoresistance measurements are done. The triangles green, yellow, and red represent a hole-doped bias, a charge neutrality bias, and an electron-doped bias respectively. 
    \textbf{b, c, d,} The longitudinal resistance $R_{\mathrm{xx}}$ and magnetoresistance $(\mathrm{MR}= [R_{\mathrm{xx}}(B)-R_{\mathrm{xx}}(B=0)]/R_{\mathrm{xx}}(B=0))$ as a function of in-plane magnetic field $B_{\mathrm{||}}$ at dopings identified by the symbols on the upper-right corner. The arrows indicate the direction of the magnetic field sweep.
    }
    }
\end{figure*}
\newpage

\begin{figure*}[h!]
    \centering
    \includegraphics{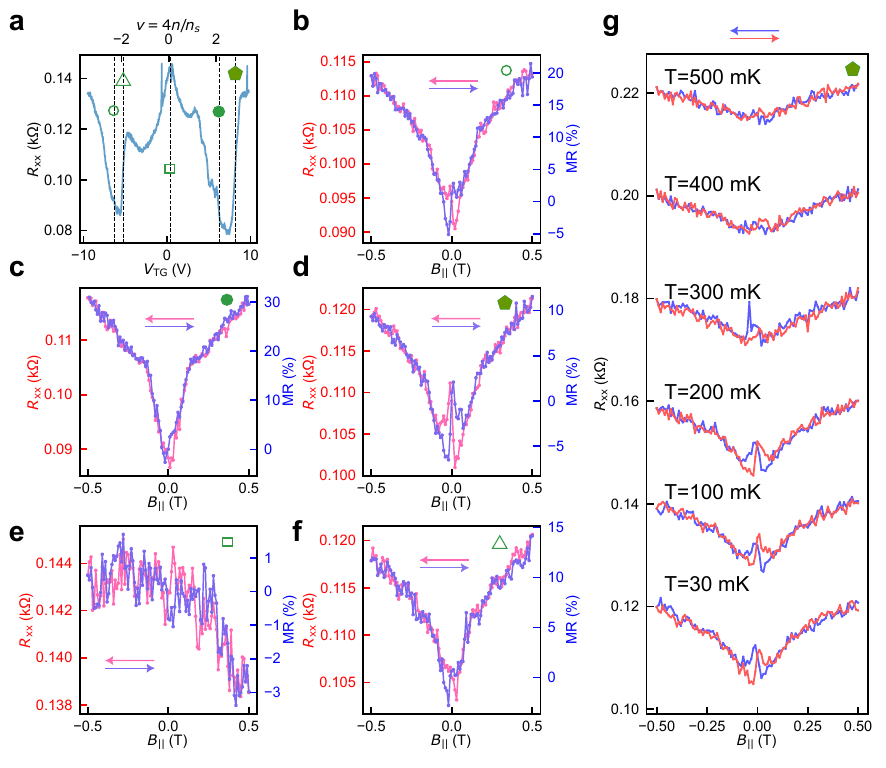}
    \caption{ \label{fig:fig8} 
    \textbf{Magnetoresistance and hysteresis in non-superconducting device with angle $1.45^{\circ}$. }
    {\textbf{a,} Lineslice of longitudinal resistance $R_{\mathrm{xx}}$ as a function of top gate voltage $V_{\mathrm{TG}}$ (the top axis indicates the filling factor $\nu$) for device C that does not show superconductivity but shows the characteristic dip in $R_{\mathrm{xx}}$, which drops to a small non-zero value. The dashed lines identify the dopings for the following magnetoresistance measurements.
    \textbf{b - f,} Longitudinal resistance $R_{\mathrm{xx}}$ and magnetoresistance $(\mathrm{MR}= [R_{\mathrm{xx}}(B)-R_{\mathrm{xx}}(B=0)]/R_{\mathrm{xx}}(B=0))$ as a function of in-plane magnetic field $B_{\mathrm{||}}$ at dopings identified by the symbols on the upper-right corner. The arrows indicate the direction of the magnetic field sweep.
    \textbf{g,} $R_{\mathrm{xx}}$ as a function of $B_{\mathrm{||}}$ for different temperatures. Similar to Fig. \ref{fig:fig3}f, the hysteresis vanishes as the temperature is increased.
    }
    }
\end{figure*}
\newpage

\begin{figure*}[h!]
    \centering
    \includegraphics{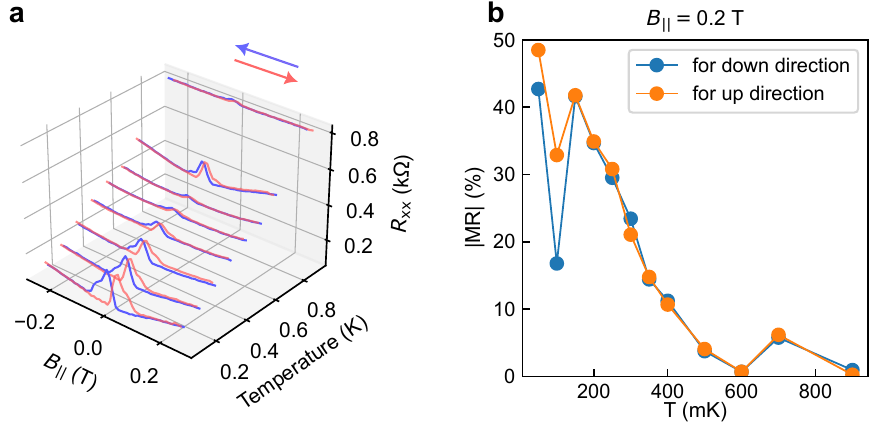}
    \caption{ \label{fig:fig9}
    \textbf{Visualisation of the hysteresis and magnetoresistance as a function of temperature.}
    {\textbf{a,} Lineslices of longitudinal resistance $R_{\mathrm{xx}}$ with in-plane magnetic field $B_{\mathrm{||}}$ at different temperatures in a 3D plot at the same doping as for Fig. \ref{fig:fig3}f. It is plotted with fewer line slices than in Fig. \ref{fig:fig3}f for clear viewing. 
    \textbf{b,} Absolute magnetoresistance $(\mathrm{MR}= |R_{\mathrm{xx}}(B)-R_{\mathrm{xx}}(B=0)|/R_{\mathrm{xx}}(B=0))$ as a function of temperature T at in-plane magnetic field $B_{\mathrm{||}}=0.2$~T for up and down direction of magnetic field sweep.
    }
    }
\end{figure*}
\newpage
 
\begin{center}
	\textbf{\Large SUPPLEMENTARY INFORMATION\\ \normalsize Superconducting magic-angle twisted trilayer graphene hosts \\competing magnetic order and moiré inhomogeneities}
\end{center}

\setcounter{section}{0}
\renewcommand{\thesection}{\Roman{section}}

\setcounter{figure}{0}
\renewcommand{\figurename}{Supplementary Fig.}
\renewcommand{\thefigure}{S\arabic{figure}}

\renewcommand{\theHequation}{Sequation.\theequation}
\renewcommand{\theequation}{S\arabic{equation}}
\setcounter{equation}{0}
\renewcommand{\thefigure}{S\arabic{figure}}

\section{I. CONTINUUM MODEL HAMILTONIAN AND DENSITY OF STATES}

In a twisted trilayer graphene, the top (\( l = 1 \)) and bottom (\( l = 3 \)) layers are aligned, while the middle layer (\( l = 2 \)) is rotated with respect to the top and bottom layers by an angle \( \theta \). For the non-interacting Hamiltonian of electrons 
we adopt the continuum model used in recent studies
\cite{C_lug_ru_2021,Xie_2021,qin2021plane,lake2021reentrant,christos2022correlated}, which extends the Bistritzer-MacDonald model of the twisted bilayer graphene (TBG) \cite{Bistritzer_2011} to the trilayer system. The Hamiltonian acting on the sublattices of the three layers of graphene \( (A_1, B_1, A_2, B_2, A_3, B_3 )\) is
given by:
\begin{align}
 \mathcal H = 
	\begin{pmatrix}
		H_0(\bm k_1) & T &  \\
		T^\dagger & H_0(\bm k_2) & T^\dagger \\
		 & T & H_0(\bm k_3)
	\end{pmatrix},
	\label{eq:ham}
\end{align}
where \( \bm k_l = \bm K_{\xi} + R\left((-1)^l\theta / 2\right) \bm k \), 
\(\bm K_\xi = \xi  (4 \pi / 3 a,0) \), \( a=2.46  ~\text{\AA}\) is the graphene lattice constant, 
\(R(\theta) = \sigma_0\cos \theta  - i \sigma_y\sin \theta  \) is the rotation
matrix, and \( H_0(\bm k) = - \hbar v_{\mathrm F} ( \xi \sigma_x, \sigma_y)
\cdot \bm k \), with Dirac velocity \( v_{\mr F} = 10^6 \) m/s, and $\sigma_{j}$ represents the Pauli matrices acting on the sublattice space. Here, \(
\xi = \pm 1 \) represent K and K$'$ valleys of graphene, respectively. \( T \)
denotes the interlayer coupling matrix that induces the moiré potential, given by: 
\begin{align}
  &T = T_0 + T_1 e^{i \xi \bm G_1 \cdot \bm r} + T_2 e^{i \xi (\bm G_1 + \bm G_2 )\cdot \bm r}, \\
  &T_n = w_{AA} \sigma_0 + w_{AB} (\sigma_x \cos( n \phi) 
+ \sigma_y \sin(n \phi)
), \quad \phi = 2\pi / 3,
\end{align}
where the reciprocal lattice vectors are 
given by
\(\bm G_1 = -k_\theta \left(\sqrt 3 / 2, 3 / 2\right),\, \bm G_2 = k_\theta\left(\sqrt 3 , 0\right)
\) with the moiré momentum scale defined by \( k_\theta = 4 \pi \sin(\theta / 2) / 3a \). Here, the interlayer hoppings follow
\( w_{AA} < w_{AB} \), which incorporates the out of plane corrugation of the layers \cite{Koshino_2018}.
For our calculation we use \( w_{AA} = 0.7 w_{AB} \) and \( w_{AB} = 124 \) meV \cite{christos2022correlated}.

The Hamiltonian \eqref{eq:ham} exhibits mirror symmetry, with the mirror
operation defined as a reflection about the middle layer, \( \mathcal M:
(1,2,3) \mapsto (3,2,1) \). In presence of a uniform out-of-plane electric field, the
Hamiltonian \eqref{eq:ham} includes an additional term
\begin{align}
	\mathcal V = 
\begin{pmatrix}
	\frac{U}{2} &  &  \\
	 & 0 &  \\
	 &  & -\frac{U}{2} \\
\end{pmatrix} \otimes \sigma_0,
\end{align}
where \( U \) is the electrostatic potential difference between the top and bottom layers. The
presence of $\mathcal V$ introduces layer asymmetry, which breaks the mirror
symmetry.
\begin{figure}
    \centering
    \includegraphics[width=0.9\linewidth]{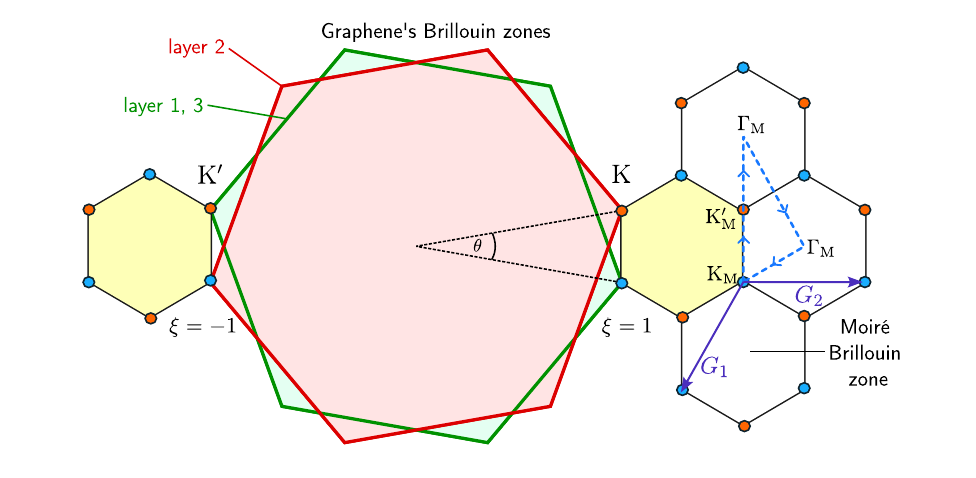}
    \caption{\textbf{Moiré Brillouin Zone.} The large green (red) hexagon depicts the Brillouin zone of graphene's layer 1,3 (layer 2), while the smaller hexagon denote the moiré Brillouin zone of the twisted trilayer system. The path illustrated, connecting high-symmetry points (\(\mathrm{K}_{\mathrm{M}} \rightarrow \mathrm{K}_{\mathrm{M}}' \rightarrow \mathrm{\Gamma}_{\mathrm{M}} \rightarrow \mathrm{\Gamma}_{\mathrm{M}} \rightarrow \mathrm{K}_{\mathrm{M}}\)), is used in Fig.~\ref{fig:linecut}.}
    \label{fig:BZ}
\end{figure}
The Hamiltonian is diagonalized through a unitary transformation into the plane
wave basis with momentum \( \bm k + \bm G \), where \( \bm k \) lies within the
moiré Brillouin zone (mBZ) and \( \bm G = n \bm G_1 + m \bm G_2 \), with \( n,m \in
\mathbb Z \) is reciprocal lattice vectors. For convergent low-energy
bands within a $100$~meV range it is sufficient to restrict \( \abs{n} , \abs{m} \leq 3 \). 
The matrix elements in the plane wave basis read:
\begin{align}
	\mathcal H_{\bm G, \bm G'} &= \begin{pmatrix}
	H_0(\bm k_1 + \bm G) + \frac{U}{2} &  &  \\
	 & H_0(\bm k_2 + \bm G) &  \\
	 &  & H_0(\bm k_3 + \bm G) - \frac{U}{2}
\end{pmatrix} \delta_{\bm G, \bm G'}
 \\ 
							   & \hspace{1em}+ \begin{pmatrix}
	0 & 0 & 0 \\
	1 & 0 & 1 \\
	0 & 0 & 0 \\
\end{pmatrix} \otimes\left(
T_0
 \delta_{\bm G, \bm G'} + 
T_1
 \delta_{\bm G, \bm G' + \xi \bm G_1} +
T_2
 \delta_{\bm G, \bm G' + \xi (\bm G_1 + \bm G_2)} 
\right) \nonumber \\
&\hspace{1em}+
 \begin{pmatrix}
	0 & 1 & 0 \\
	0 & 0 & 0 \\
	0 & 1 & 0 \\
\end{pmatrix} \otimes\left(
T_0
 \delta_{\bm G, \bm G'} + 
T_1
 \delta_{\bm G + \xi \bm G_1, \bm G'} +
T_2
 \delta_{\bm G + \xi (\bm G_1 + \bm G_2), \bm G' } 
\right). \nonumber
\end{align}
The resultant low-energy dispersion is shown in Fig. \ref{fig:linecut} for a range
of twist angles. In the absence of an electric field, the low-energy bands resemble
TBG-like flat bands and Dirac cones (see Fig.~\ref{fig:linecut} \textbf{a}-\textbf{f}), each belonging to distinct mirror sectors.
However, in the presence of an electric field, these mirror sectors hybridize,
causing a mixing of the TBG-like bands and Dirac cones (see Fig.~\ref{fig:linecut} \textbf{g}-\textbf{l}).

We also incorporate a valley-exchange term in the Hamiltonian as follows:
\begin{align}
\tilde{\mathcal H} = 
\begin{pmatrix}
	\mathcal H^{(\xi = +1)} & \Delta_{\mr{V}} \mathbb I \\
	\Delta_{\mr{V}} \mathbb I & \mathcal H^{(\xi = -1)}
\end{pmatrix}. \label{eq:valley-ex}
\end{align}
Such valley-exchange can be generated spontaneously from Coulomb interaction \cite{C_lug_ru_2021,christos2022correlated}. The presence of valley-exchange term breaks valley $U(1)_{\mathrm V}$ symmetry and gives rise to intervalley coherence. In general $\Delta_{\mr{V}}$ depends on momentum, however, in the following we consider $\Delta_{\mr{V}}$ to be independent of momentum for simplicity.
\begin{figure}[ht]
	\centering	\includegraphics[width=\textwidth]{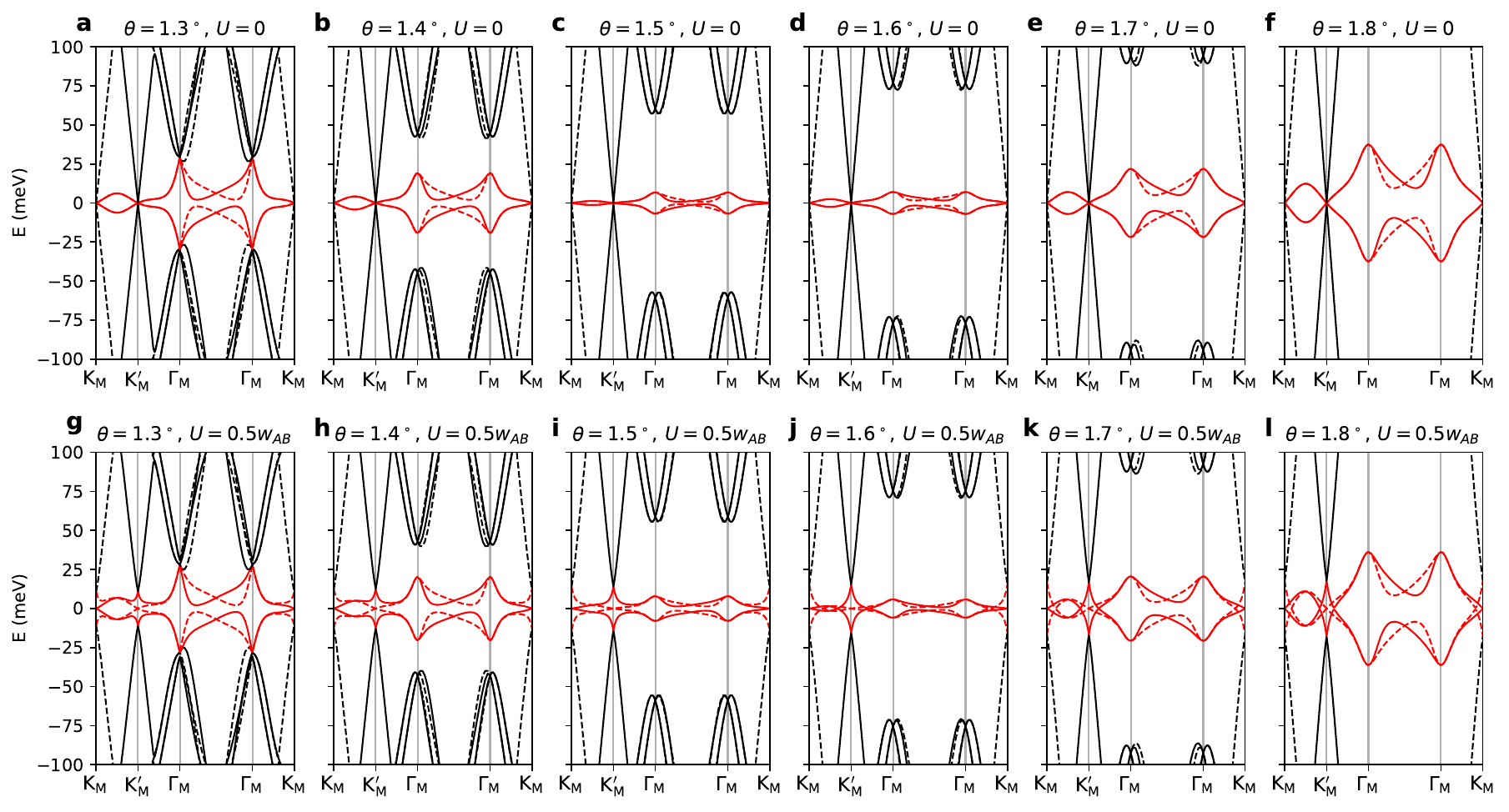}
	\caption{\textbf{Electronic band structure of twisted trilayer graphene.} Band structure of the continuum Hamiltonian (Eq.~\eqref{eq:ham}) plotted along high-symmetry points for both valleys of graphene. Solid (dashed) lines represent the K$'$ (K) valley of graphene, with angles ranging from $\theta =  1.3^\circ - 1.8^\circ$. The flat bands are highlighted in red. Parameters used 
 here: \( w_{AB} = 124 \) meV, \( w_{AA} = 0.7 w_{AB} \), 
 \textbf{a}-\textbf{f}: For $U=0$, mirror symmetry is preserved, and TBG-like bands and Dirac bands belonging to different mirror-symmetry sectors do not hybridize at any angles.
 \textbf{g}-\textbf{l}: At $U=0.5w_{AB}$, the electrostatic potential induces layer asymmetry, breaking mirror symmetry. Consequently, TBG-like bands and Dirac bands hybridize.}
	\label{fig:linecut}
\end{figure}

The carrier density corresponding to a given chemical potential $\mu$ can be determined from fermion number conservation, as described by the equation:
\begin{align}
	\label{eq:number-equation}
	n_0 + n = \frac{1}{V}\sum_{\bm k \in \mr{mBZ}} \sum_i f(\epsilon_i(\bm k) - \mu).
\end{align}
Here, \( n_0 \) represents the total density up to the charge neutrality, while
\( n \) is the carrier density relative to the charge-neutrality, and \(\epsilon_i(\bm k)\) are
the energies.
\(
f(\cdot) \) represents the Fermi-Dirac distribution. The density
corresponding to a completely filled band is \( n_s  = 4 / \Omega\) ($\Omega$
is the area of a moiré unit cell), which corresponds to a filling four electrons
in a moiré unit cell. Using this we can define the filling factor \( \nu =
4 n / n_s \). Filling the flat bands corresponds to \( \nu \) within a range of  $[-4,4] $. We use Eq.~\eqref{eq:number-equation} to find
relation between \( \mu \) and \( \nu \), below we compute the density of
states (DOS) as a function of \( \nu \). This allows for a comparison of the
density of states at different angles on an equal footing.
\begin{align}
	\mathrm{DOS}(\nu) = \frac{1}{V}\sum_{\bm k \in \mr{mBZ}, i} \delta(\epsilon_i(\bm k) - \mu(\nu)).
\end{align}
Here \( V = N \Omega \), where \( N \) is the total number of unit cells. For
calculation of DOS we use \( N = 60 \times 60 \) and the Lorentzian representation of the Dirac delta function with a width of 0.1 meV.

\begin{figure}
    \centering
\includegraphics{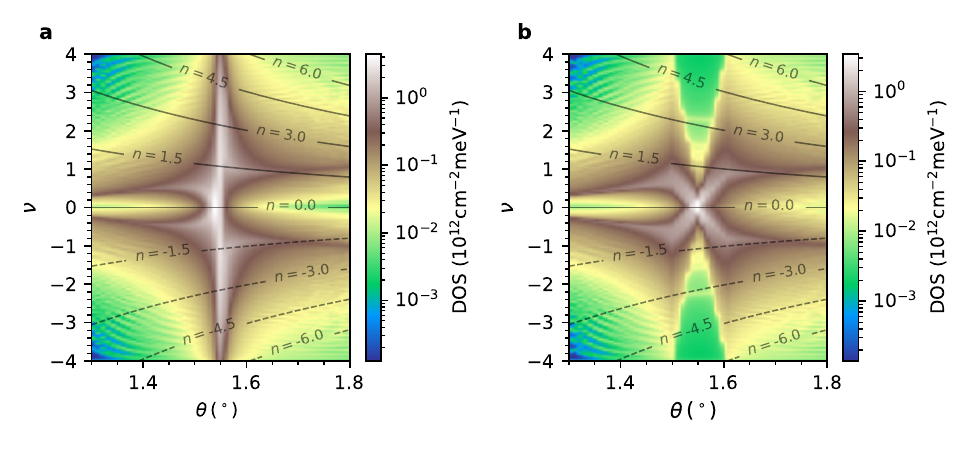}
   \caption{\textbf{ Twist-angle and filling factor dependence of density of states.} Density of states plotted as a function of twist-angle ($\theta$) and filling factor ($\nu$) for the given parameters:
   $w_{AB} = 124~ \mr{meV}, \, w_{AA}=0.7 w_{AB}$.
   The contour lines represent the carrier density $n$ ($10^{12}~\mathrm{cm}^{-2}$) and show how, given a certain carrier density, the filling factor varies with twist angle. 
	\textbf{a}, For $U = 0$ the density of states exhibits a sharp peak at the magic angle ($\theta \approx 1.56^\circ$).
 \textbf{b},~For $U = 0.25 w_{AB}$ there are significant changes in the density of states, which is no longer sharply peaked at the magic angle, indicating rapid variations of the bands with the electrostatic field.
 }
	\label{fig:DOS2}
\end{figure}

\begin{figure}
    \centering
\includegraphics{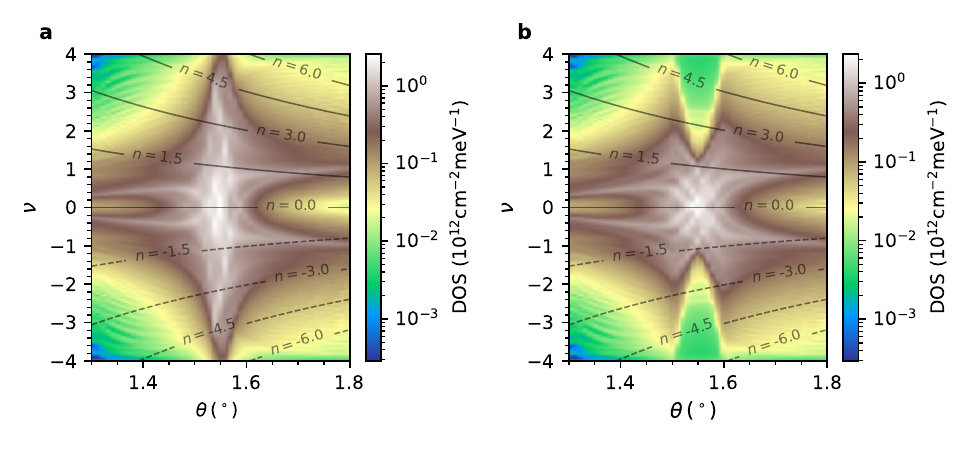}
	\caption{\textbf{Effect of valley exchange.} 
Density of states plotted as a function of twist-angle ($\theta$) and filling factor ($\nu$) for the given parameters:
   $w_{AB} = 124~ \mr{meV}, \, w_{AA}=0.7 w_{AB}$ with valley exchange term $\Delta_{\mr{V}} = 10~ \mr{meV}$. The contour lines represent the carrier density $n$ ($10^{12}~\mathrm{cm}^{-2}$) and show how, given a certain carrier density, the filling factor varies with twist angle.
	\textbf{a}, For $U = 0$ and
 \textbf{b}, for $U = 0.25 w_{AB}$.
 The distinct features of the density of states broaden, yet remain qualitatively similar to those in Fig. \ref{fig:DOS2}.
}
	\label{fig:DOS3}
\end{figure}

We show the density of states for a continuous range of twist
angle without (Eq.~\eqref{eq:ham}) and with the valley-exchange term (Eq.~\eqref{eq:valley-ex}) in Fig.~\ref{fig:DOS2} and Fig.~\ref{fig:DOS3}, respectively.
For $U=0$, we observe sharp density of states around the magic angle $1.56^\circ$.
As the bandwidth is minimum at magic-angle (see Fig. \ref{fig:linecut}), leading to a large density of states, one expects correlation-driven instabilities
to become more likely. Conversely, away from the magic angle, correlation effects are expected to be relatively weak. The density of states change rapidly with an applied electric field, which no longer peak at the magic angle as the filling factor deviates from zero.

We superimpose the density of states with contour lines representing a fixed carrier density (see Fig.~\ref{fig:DOS2}, \ref{fig:DOS3}) to assess variations across different regions of the twisted trilayer system. STM observations reveal the presence of plaquette (P), soliton (S), and twiston (T) regions due to lattice relaxation \cite{turkel2022orderly}, each characterized by local twist angles obeying $\theta_\mathrm{P} (\approx 1.56^\circ) < \theta_\mathrm{S} < \theta_\mathrm{T}$. Consequently, if we maintain a constant carrier density, the local filling factors obey $\nu_\mathrm{P} > \nu_\mathrm{S} > \nu_\mathrm{T}$. Thus, while the plaquette region resides within the superconducting phase, the soliton and twiston regions exhibit lower local filling factors.

\newpage
\section{II. CHARACTERIZATION OF TWIST ANGLE INHOMOGENEITY IN THE MATTG DEVICES}

\begin{figure}[h!]
    \centering
    \includegraphics[width=\textwidth]{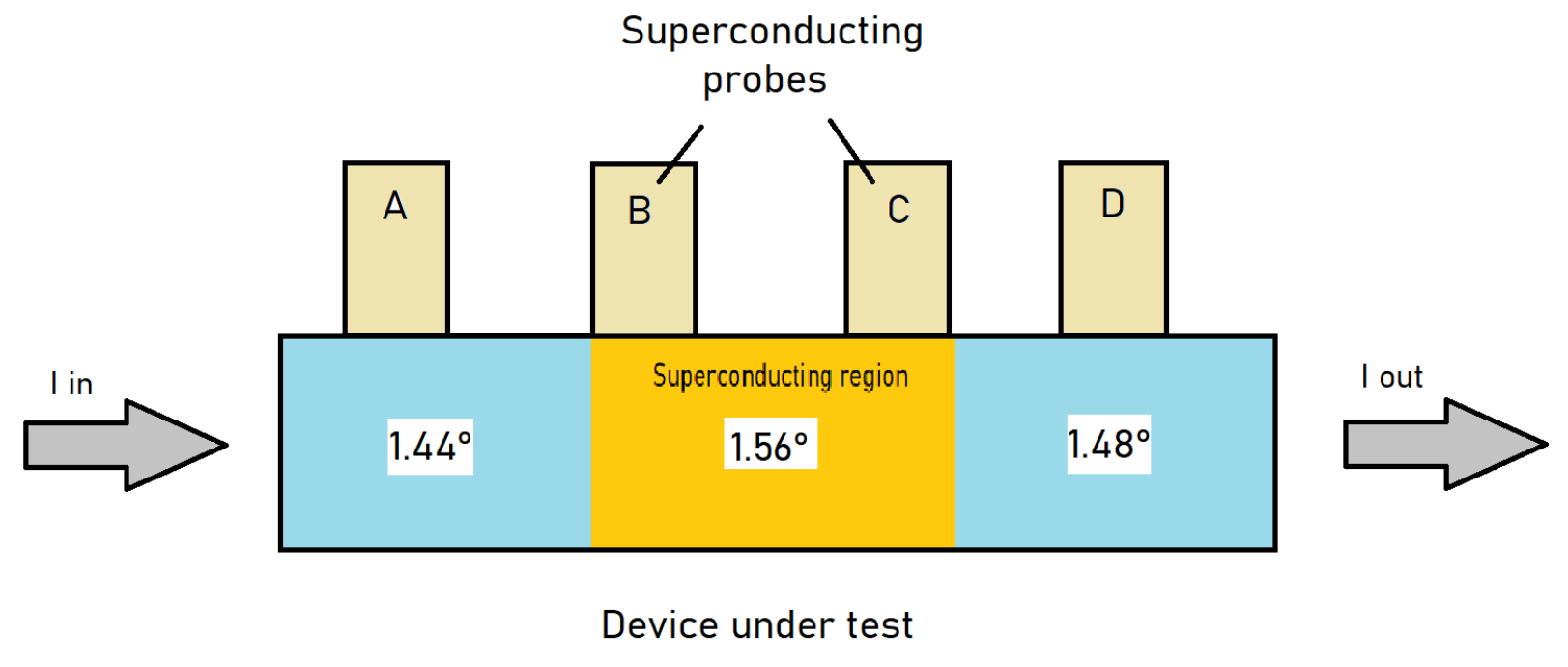}
    \caption{\label{fig:figS5}\textbf{Schematic showing the distribution of twist angle in device - A due to twist angle disorder.}
     Schematic showing the different probes in device - A and the angles inferred from measurements in these adjacent probe combinations in the device under test. The probes B-C turn out to be superconducting, while the others do not.  
    }
    \end{figure}

Twist angle disorder is ubiquitous in all moir\'e twisted devices. This disorder can arise primarily from strain. However, such a disorder is distinct from the relaxation-induced moir\'e inhomogeneities that give rise to moir\'e solitons and twistons. 

Suffice it to say, that it still is important to characterize devices based on the twist angle disorder. Due to such disorders, we only find some pairs of probes in devices to show superconductivity and not all. Fig.~\ref{fig:figS5} shows the different twist angles inferred from transport data in the various probes of device - A. Table~\ref{table:table1} lists the twist angles in device~-~B.

\begin{table}[h!]
\centering
 \begin{tabular}{||c| c| c||} 
 \hline
 Probe pair & Angle inferred from transport & Superconductivity shown \\ [0.5ex] 
 \hline\hline
 Pair 1: A-B & 1.43$^{\circ}$ & Yes  \\ 
 \hline
 Pair 2: B-C & 1.39$^{\circ}$ & No  \\
 \hline
 Pair 3: C-D & 1.37$^{\circ}$ & No \\
 \hline
 Pair 4: D-E & 1.36$^{\circ}$ & No \\
[1ex] 
 \hline
 \end{tabular}
 \caption{Table showing the distribution of angles in consecutive probes in device - B (Fig.~\ref{fig:figS8}) to characterize the twist angle disorder across the device length.}
 \label{table:table1}
\end{table}

\newpage
\section{\label{sec:S3} III. SAMPLE PREPARATION}

\begin{figure}[h!]
    \centering
    \includegraphics{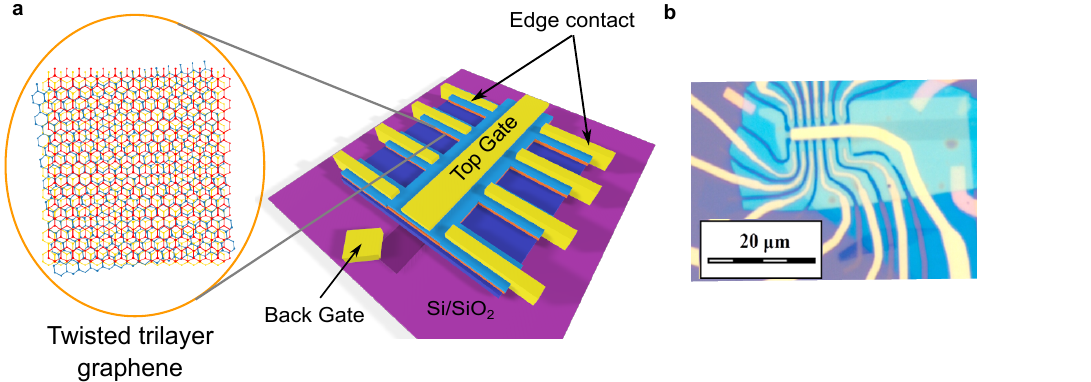}
    \caption{\footnotesize \label{fig:figS6} {\textbf{Device schematic and optical image.}
    \textbf{a,}~Schematic of the device. Top gate (gold) voltage $V_{\mathrm{TG}}$ and bottom gate (graphite) voltage $V_{\mathrm{BG}}$ is applied to control the carrier density $n$ and hence, filling $\nu$ and perpendicular electric displacement field $D$.
    \textbf{b,}~Optical image of the device A. The contact electrodes are edge-contacted to the twisted trilayer graphene.
    }}
\end{figure}

The magic-angle twisted trilayer graphene (MATTG) is made by stacking three layers of graphene with the top and bottom layers aligned, and the middle layer twisted by an angle close to the magic angle. These three layers are then encapsulated by a top and a bottom hexagonal boron nitride (hBN) of around 20-40~nm thickness. Graphene and hBN are exfoliated on p-doped Si/SiO$_{2}$ substrate, and their thickness is determined optically.  In order that the angle between the three graphene layers of the stack is maintained, all pickups are done using a motorized stage. Each of these three layers is picked up from the same monolayer graphene flake by cutting it into three pieces using a tapered fiber optic scalpel \cite{sangani2020facile}. The stacking is done using a dry-pickup technique using a PC/PDMS layer placed on a glass slide. The hBN flakes are picked up at a temperature of 90~$^{\circ}$C, while the graphene flakes are picked up at a temperature of 70~$^{\circ}$C. A half stack consisting of the top hBN and three graphene flakes is dropped on an already dropped bottom stack of hBN and graphite gate. It is then vacuum annealed at 350~$^{\circ}$C \cite{zhao2012quantum}. The final device lies on an intrinsic Si/SiO$_{2}$ substrate. The top gate and bottom gate are made using e-beam lithography and e-beam evaporation of Cr/Au. The edge contacts to graphene are made by e-beam lithography, CHF$_{3}$ reactive ion etching, and e-beam evaporation of Cr/Pd/Au. Fig.~\ref{fig:figS6} shows a schematic and optical image of the device.

\section{IV. MEASUREMENT TECHNIQUE} \label{sec:S4}

All measurements are done in a dilution refrigerator with a base temperature of 20~mK. The voltage drops across the device are amplified by a factor of 1000 by DL Instruments pre-amplifier, before measuring them using an SR830 lock-in amplifier for AC measurements and an NI-DAQ for DC measurements. A current of around 5~nA is used for the differential resistance measurements. All lock-in amplifiers used are synced to a frequency of around 17~Hz. The top and the bottom gate voltages are applied using an NI-DAQ, and are filtered using an low pass 10~Hz RC filter. There are three additional stages of RC filtering in the dilution refrigerator – at room temperature, at 4~K plate, and at the 20~mK plate – all 70~kHz low pass filters to protect the device from high-frequency noise. There are also passive filters in the measurement line like copper powder filters, at 20~mK and ECOSORB filters at 4~K, to attenuate other higher frequency noise.

\section{V. CRITICAL TEMPERATURE OF MATTG SUPERCONDUCTOR}

\begin{figure}[h!]
    \centering
    \includegraphics[width=17.2cm]{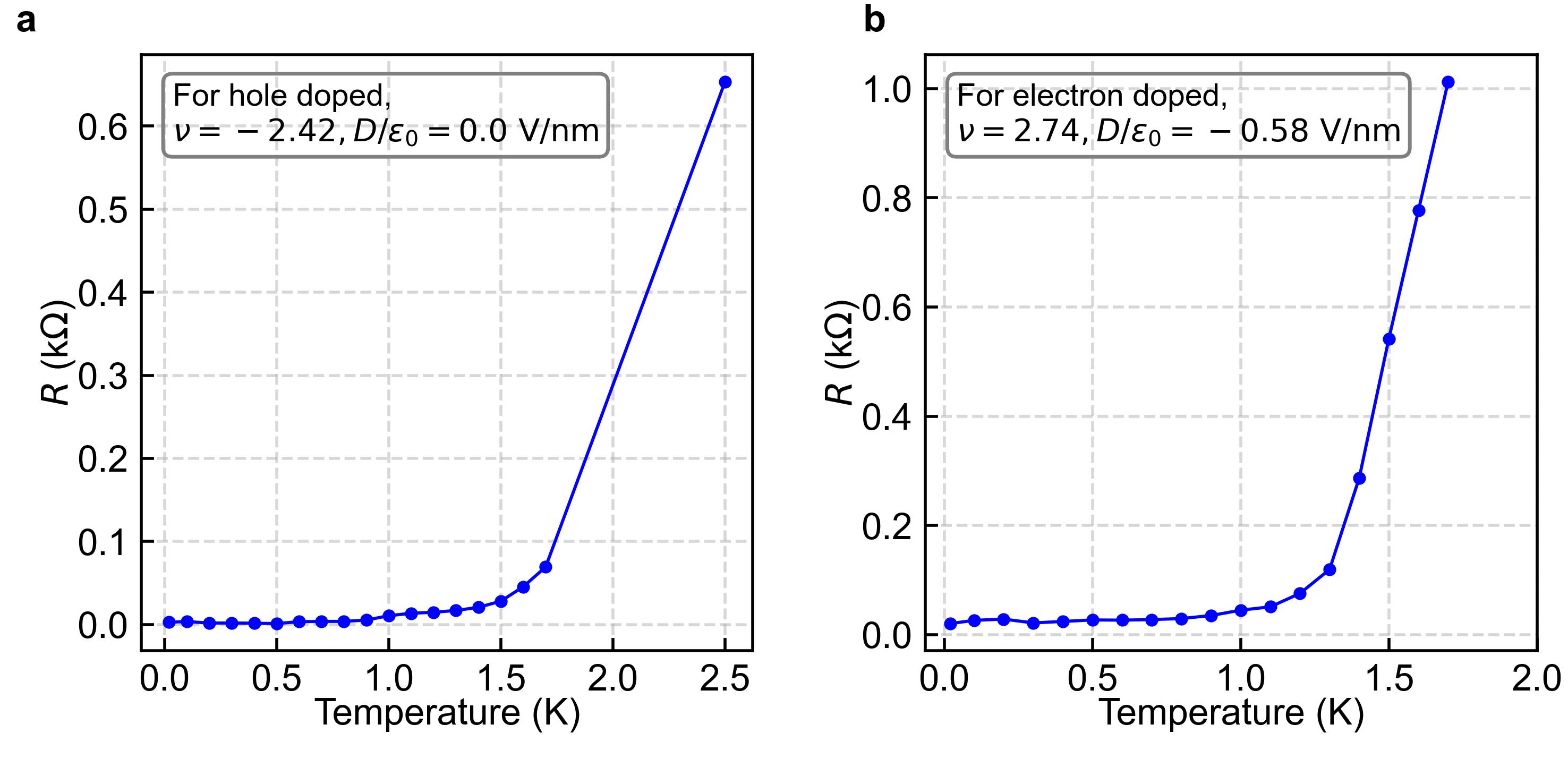}
    \caption{\label{fig:figS7}\textbf{R vs T data for hole and electron side superconductivity.}
    \textbf{a,}~R vs T for the optimal hole doping of $\nu=-2.42,~D/\epsilon_0=0.0~$V/nm. 
    \textbf{b,}~R vs T for the optimal electron doping of $\nu=2.74,~D/\epsilon_0=-0.58~$V/nm. Both data are consistent with existing literature.
    }
   \end{figure}

The R vs T plots for the superconducting phase of both the optimal hole-doped and optimal electron-doped regions are shown in Fig.~\ref{fig:figS7}. The optimal doping and electric field is chosen at the regions showing the maximum critical current $I\mathrm{_{c}}$. The jump of the resistance from zero to a non-zero value guides us toward the value of the critical temperature $T\mathrm{_{c}}$.

\section{VI. ANOTHER SUPERCONDUCTING DEVICE}

\begin{figure}[h!]
    \centering
    \includegraphics{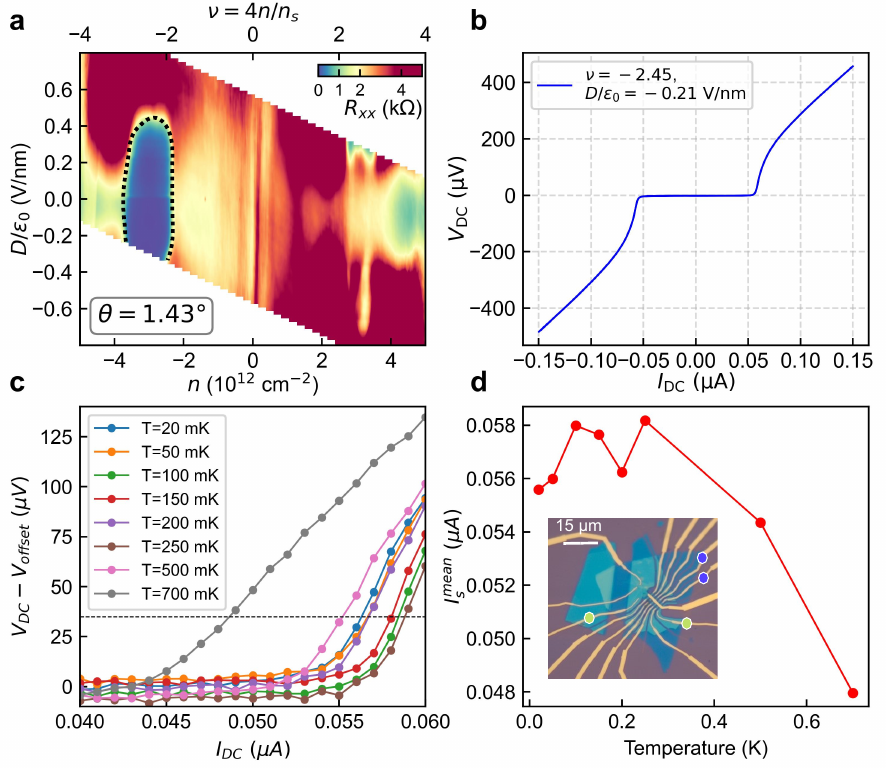}
   \caption{\label{fig:figS8}\textbf{Characteristics of another superconducting device - device B.} 
   \textbf{a,} Longitudinal resistance $R_{\mathrm{xx}}$ plotted as a function of carrier density $n$, filling $\nu$ and electric field $D/\epsilon_{o}$. The superconducting zero resistance region of the device is marked by a black dashed line.
   \textbf{b,} DC I-V characteristic of the device at the optimum filling and electric field. A critical current of $\sim$55~nA is hosted by the device at $T$=20~mK.
   \textbf{c,} DC I-V characteristics zoomed in near the transition current and plotted as a function of temperature $T$. A voltage offset $V_{\mathrm{offset}}=15.915~\mathrm{\mu V}$ is subtracted from the DC voltage $V_{\mathrm{DC}}$ to correct for offsets in the measurement setup.
   \textbf{d,} The mean switching current extracted from 1000 DC I-Vs at each temperature, marking the current encountered just before $V_{\mathrm{DC}}$ crosses 35~$\mathrm{\mu V}$. The non-monotonic trend of the mean switching current is evident in device B. Inset, an optical image of the device. The scale bar is 15~$\mathrm{\mu m}$. The current and the voltage probes are marked with green and blue dots, respectively.
   }
   \end{figure}

Another superconducting MATTG device - device B with twist angle $\theta=1.43^{\circ}$, was measured and its characteristics are shown in Fig.~\ref{fig:figS8}. Device B only showed superconductivity in the hole-doped regime and the critical current is $\approx 55~\mathrm{nA}$.

\section{VII. ADDITIONAL DEVICE CHARACTERIZATION}
\begin{figure}[h!]
    \centering
    \includegraphics{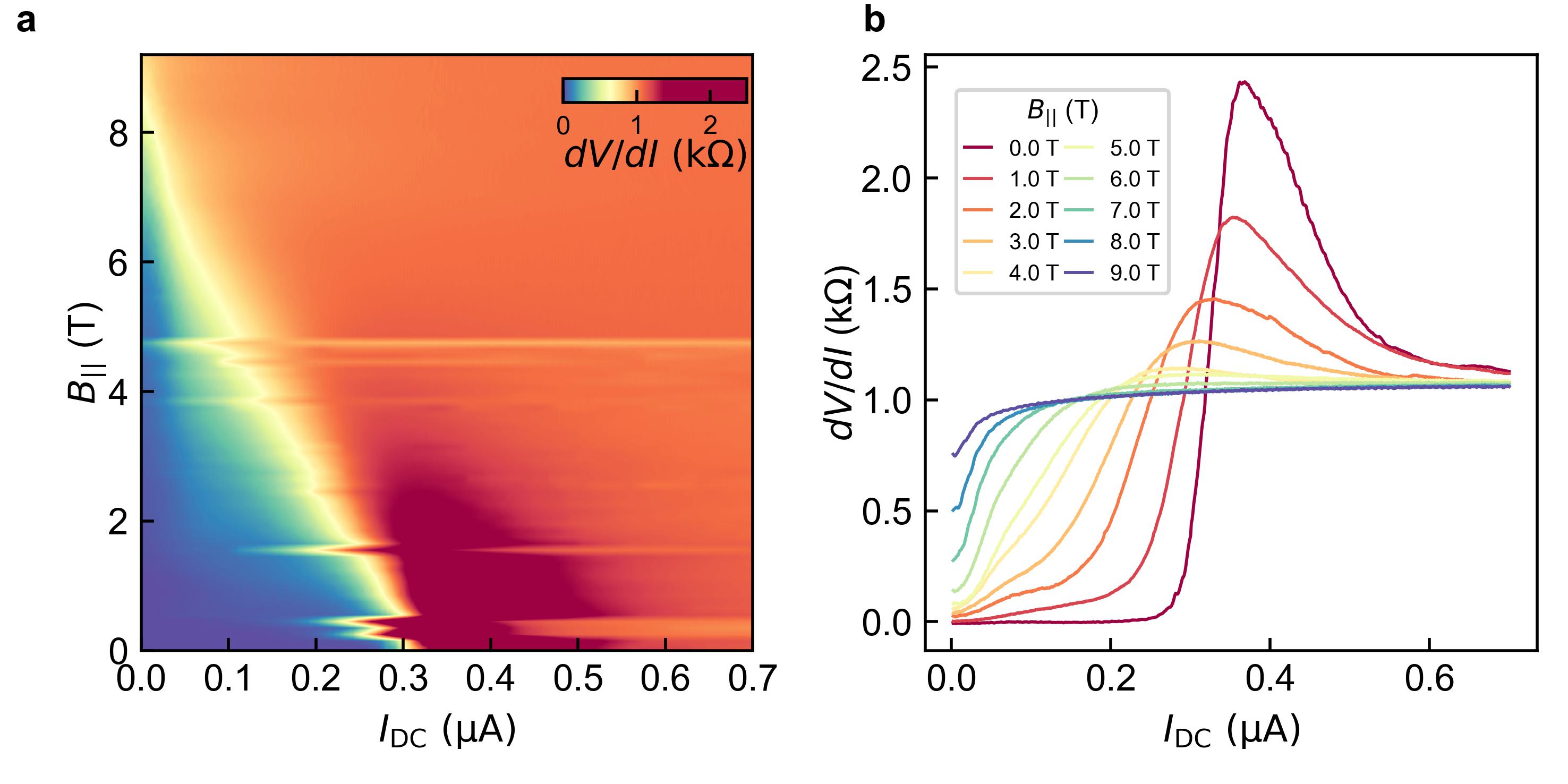}
    \caption{\footnotesize \label{fig:figS9} {\textbf{Further characterization of superconductivity in MATTG at T=20 mK.}
    \textbf{a,}~2D color scale plot of the differential resistance $dV/dI$ vs $I\mathrm{_{DC}}$ plotted as a function of in-plane magnetic field $B_{\parallel}$.
    \textbf{b,}~$dV/dI$ vs $I\mathrm{_{DC}}$ lineslices at different $B_{\parallel}$ from (a).
    }}
\end{figure}

The persistence of the superconducting phase till high values of in-plane magnetic field unlike conventional spin-singlet superconductors - as reported in MATTG \cite{cao2021pauli} is also noted in our device (see Fig.~\ref{fig:figS9}). This is attributed to Pauli-limit violation and helps to further characterize the superconductivity in MATTG.

\section{VIII. SWITCHING MEASUREMENT}
\begin{figure}[h!]
    \centering
    \includegraphics[width=17.2cm]{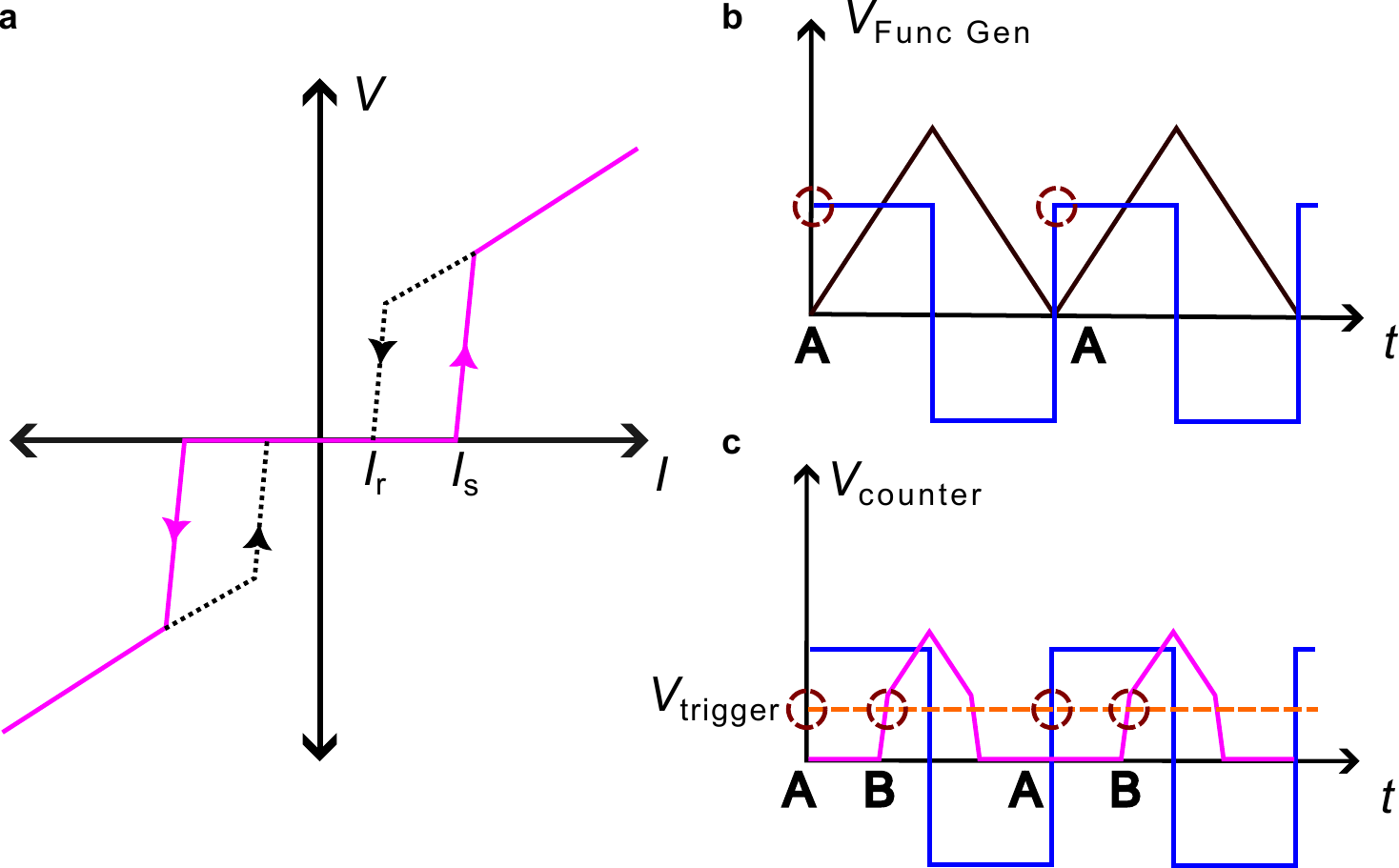}
    \caption{\footnotesize \label{fig:figS10} \textbf{Switching measurement technique.}
    \textbf{a,}~The DC I-V curve for a superconductor showing the switching current $I_{s}$ for the transition from superconductor to normal when the current increases. The retrapping current $I_{r}$ is due to the normal to superconductor transition that happens when we are decreasing the current bias in the system. The same arguments are valid for the absolute values of negative current biases.
    \textbf{b,}~A triangular wave of 10 Hz is applied using a function generator. The current varies linearly with time between 0 and a current value set slightly above the critical current. This is applied to the device. A square wave, also of 10 Hz, is applied to the counter - in blue. This signal is used to activate the counter to start its counting. Both these signals are synchronized.
    \textbf{c,}~The counter is triggered when the device voltage signal (in pink) crosses a certain threshold voltage (in orange), and it records the time elapsed between events A and B. This time then is used to get back to the switching current with the knowledge of the frequency and amplitude of the triangular signal.
    }
    \end{figure}

The switching measurements use a Tektronix FCA3100 Timer/Counter/Analyzer and Agilent 33500B series waveform generator, in addition to the setup used for measuring voltage drops in the device discussed in Sec.~IV. The counter is capable of measuring small time intervals between events defined within it - usually, the input signal crossing a certain trigger voltage. A triangular wave signal is generated using the function generator and fed through a 10 M$\Omega$ resistor acts as the current bias to the device. The current varies between zero and a maximum current of a slightly larger value than the typical switching currents. A second square signal is generated in sync by the function generator and is directly fed into port 1 of the counter. The instance of the square signal crossing zero voltage is defined as event A (see Fig.~\ref{fig:figS10}). The occurrence of event A triggers the counter to start the timer. Both signals have a frequency of 10~Hz. Port 2 of the counter receives the amplified voltage drop from the device and is set to trigger when event B occurs. Event B is defined as when the voltage drop from the device crosses $50~\mathrm{\mu}$V (that is when the DC resistance is 10\% of the normal state resistance). The counter thus records the time intervals between when the current was zero and when the device switched to the normal state - which in turn gives us a measure of the switching current, $I_{s}$~(in $\mathrm{\mu}$A) $=\frac{2\times f \times I_{\mathrm{max}}}{1000}\times t~$, where $f$ is the frequency of the signal in Hz, $I_{\mathrm{max}}$ is the maximum current bias applied to the device in $\mathrm{\mu}$A, and $t$ is the time interval measured by the counter in ms. The counter has a digital filter and a low pass filter which are set to 15 Hz and 100 Hz cut-off frequencies respectively. This mitigates spurious counter triggers due to noise.

Each of these switching measurements is carried out 10,000 times to give us large statistics of data to be analyzed. This large set of data is then represented as a histogram, with a fixed number of bins - 100. The histogram count is normalized by the total count of events and the bin width.

\section{IX. PHENOMENOLOGICAL MODEL TO UNDERSTAND NON-MONOTONIC RESPONSE}

Here we have used a simple-minded phenomenological picture to explain how the presence of a competing order gives rise to the non-monotonicity observed in the switching current $I_{\mathrm{s}}$ of the MATTG system with temperature T. The non-monotonic nature arises from a superconducting energy gap $\Delta$ competing with another order. The gap $\Delta$ increases with the increase in temperature, while the competing order weakens. Eventually, $\Delta$ reaches a maxima and falls following a Bardeen–Cooper–Schrieffer (BCS) theory picture. 

\begin{figure}[h!]
    \centering
    \includegraphics{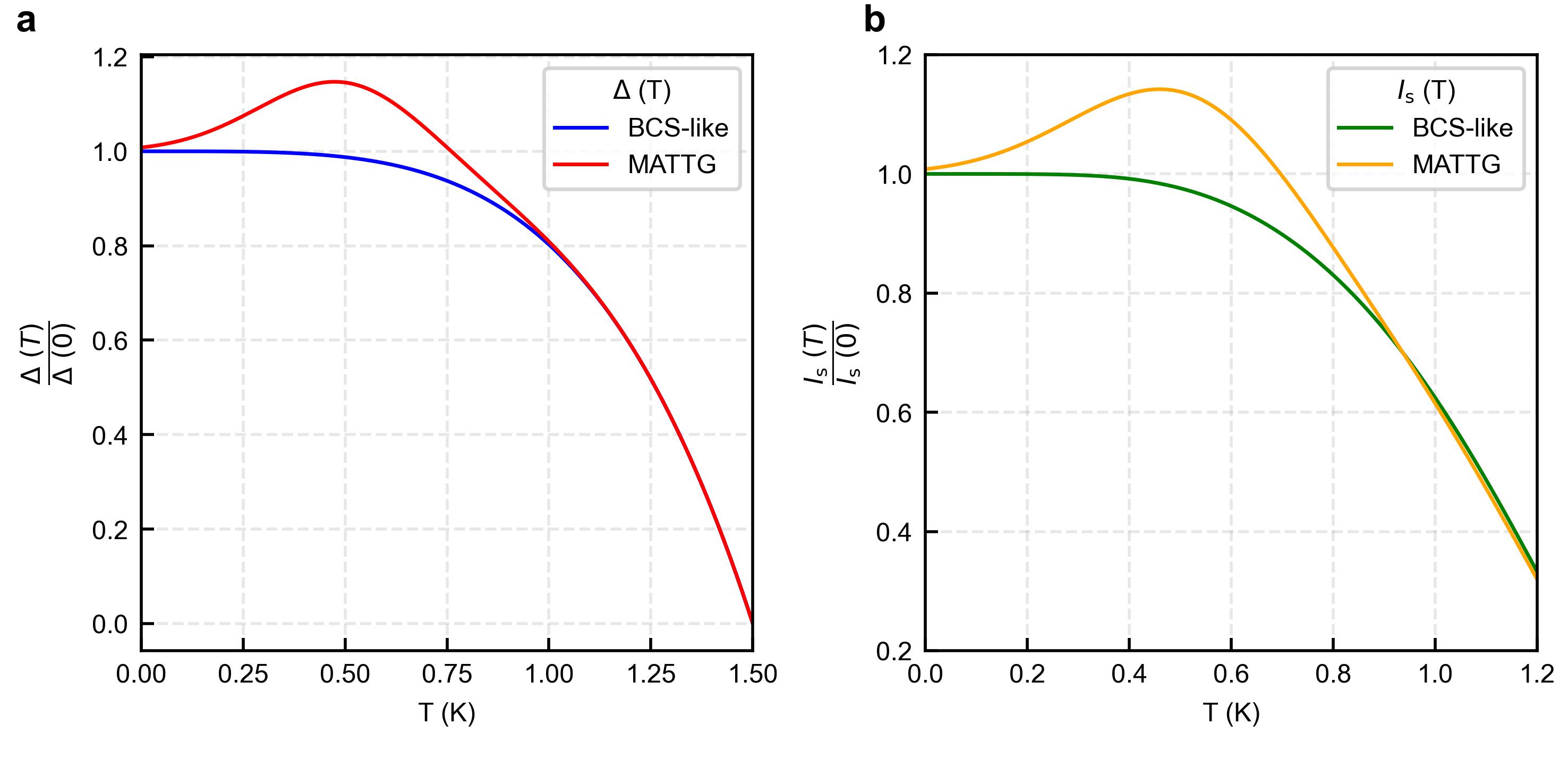}
    \caption{\footnotesize \label{fig:figS11} {\textbf{Phenomenology of a competing order}
    \textbf{a,}~Superconducting energy gap $\Delta$ as a function of temperature T for a BCS-like superconducting order and for superconducting MATTG with a competing order.
    \textbf{b,}~The switching currents $I_{\mathrm{s}}$ for the two scenarios of a JJ with a BCS-like $\Delta$ and for the MATTG JJ network. The non-monotonicity of $I_{\mathrm{s}}$ with temperature arises due to a competing order in MATTG.
    }}
\end{figure}

Fig. \ref{fig:figS11}a. shows the regular BCS-like $\Delta$ as a function of temperature and $\Delta$ (T) in MATTG having a competing phase. 
Fig. \ref{fig:figS11}b shows the $I_{\mathrm{s}}$ (T) of a JJ made of BCS superconductors and the switching current of JJ network in MATTG. This simple phenomenological model captures the non-monotonicity of $I_{\mathrm{s}}$ due to the presence of a competing order.

\section{X. ADDITIONAL SWITCHING MEASUREMENTS}
\subsection{A. Switching measurements at doping in the electron-side superconducting phase}
\begin{figure}[h!]
    \centering
    \includegraphics{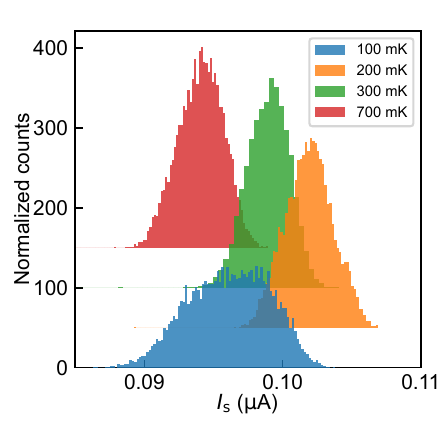}
    \caption{\label{fig:figS12}\textbf{Additional switching data at the electron-side superconducting phase.}
    The switching histograms for an electron-doped region $\nu = 3.09$, $D/\epsilon_0=0.22$~V/nm, also shows a non-monotonicity in the mean switching current of the distribution with temperature. Each histogram is shifted by 50 counts on the vertical axis for clarity.    
    }
    \end{figure}

The non-monotonic behavior of the switching current noted in Fig. 2d of the main manuscript is demonstrated for the superconductivity in the optimally hole-doped superconductor in the device. This non-monotonicity is also noted in the electron side superconductivity. It suggests that the non-monotonicity is characteristic of the superconducting phases hosted by MATTG, both hole-doped and electron-doped. The plot for the switching distributions as a function of temperature for the electron-doped SC phase is given in Fig.~\ref{fig:figS12}. Additionally, we checked the temperature calibration of the dilution refrigerator by loading a temperature sensor. 

\subsection{B. Switching measurements at additional dopings in the hole-side superconducting phase}

\begin{figure}[h!]
    \centering
    \includegraphics{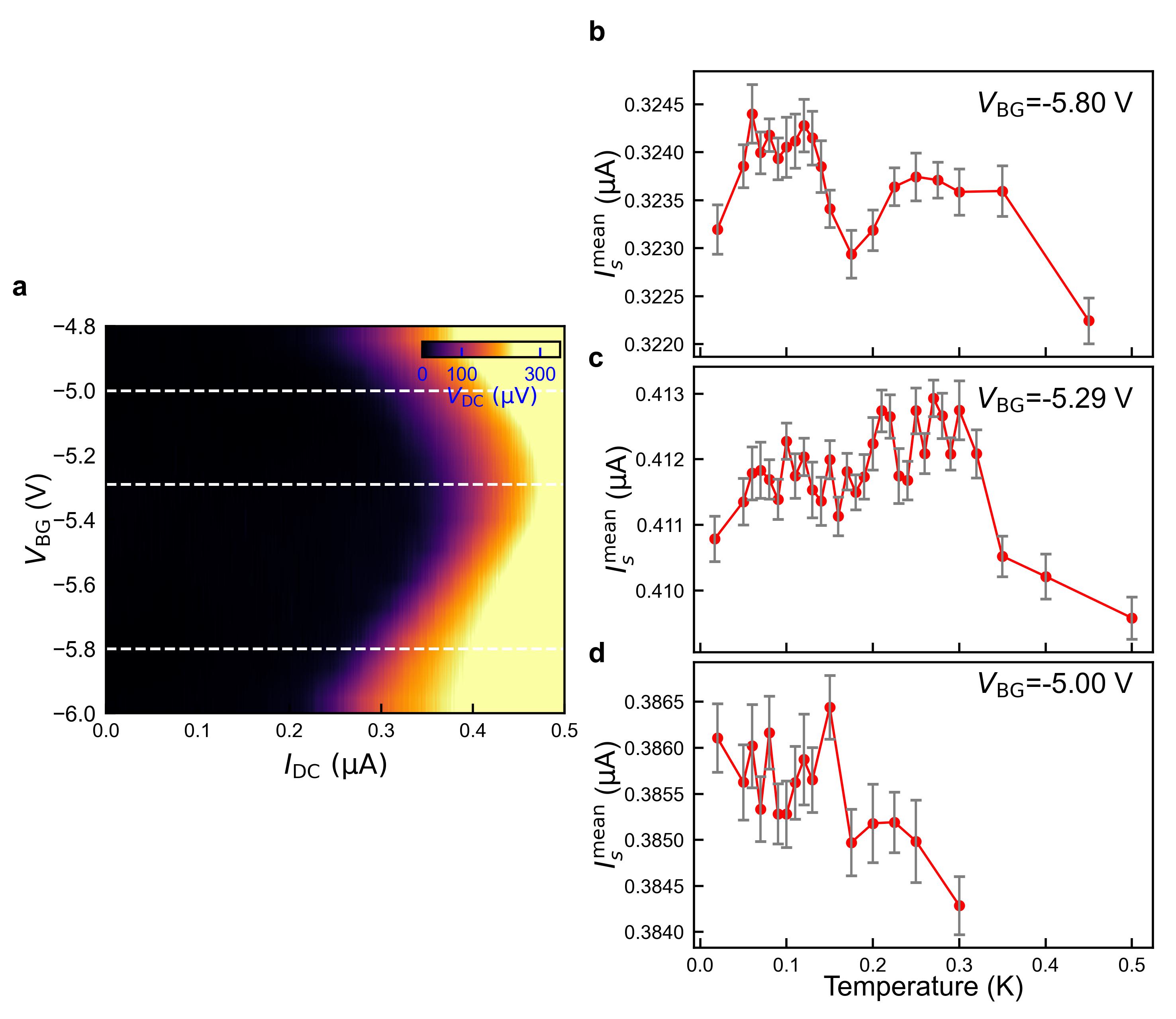}
   \caption{\label{fig:figS13}\textbf{Additional data showing evolution of mean switching current with temperature for different dopings within the hole-side superconducting phase. }
   \textbf{a,}~Voltage drop $V_{\mathrm{DC}}$ across the device measured as a function of DC current bias $I_{\mathrm{DC}}$ and bottom gate voltage $V_{\mathrm{BG}}$; parked at the hole-side superconducting phase. The top gate voltage is fixed at $V_{\mathrm{TG}} = 0 ~\mathrm{V}$. The horizontal dashed lines mark the $V_{\mathrm{BG}}$ values where the switching measurements in (b), (c), and (d) are taken.
\textbf{b, c, d,}~The mean switching current $I_{\mathrm{s}}^{mean}$ plotted as a function of temperature to observe the non-monotonic evolution of $I_{\mathrm{s}}^{mean}$ at different $V_{\mathrm{BG}}$ values; (b), $V_{\mathrm{BG}} = -5.80~ \mathrm{V}$ (c), $V_{\mathrm{BG}} = -5.29~ \mathrm{V}$ and (d), $V_{\mathrm{BG}} = -5.00~ \mathrm{V}$. The error bars represent the standard deviation of each individual histogram.
}
    \end{figure}

The evolution of the switching distribution and in particular its mean switching current $I_{\mathrm{s}}^{\mathrm{mean}}$ is studied with temperature. The evolution with temperature is then studied with different dopings (identified with horizontal lines in Fig.~\ref{fig:figS13}\textcolor{blue}{a}) in the hole-side superconducting phase that host different switching currents. With the change in the value of the switching current itself, the temperature at which the non-monotonic behavior of the $I_{\mathrm{s}}^{mean}$ is observed also changes and can be noted from Fig.~\ref{fig:figS13}\textcolor{blue}{b-d}. Biasing the device at different doping and perpendicular electric field changes the band structure in the system due to hybridization between different bands. Noting the non-monotonicity of $I_{\mathrm{s}}^{mean}$ with temperature for different points in $n$-$D$ space within the superconducting phase could reveal additional information about the competing orders involved and how they change with changing bandstructure and doping. Such a characterization with a changing perpendicular electric field at a fixed optimal doping in the hole-side superconducting phase is presented in section XC.

\newpage
\subsection{C. Switching measurements at different perpendicular electric field}\label{sec:secXIC}

\begin{figure}[h!]
    \centering
    \includegraphics{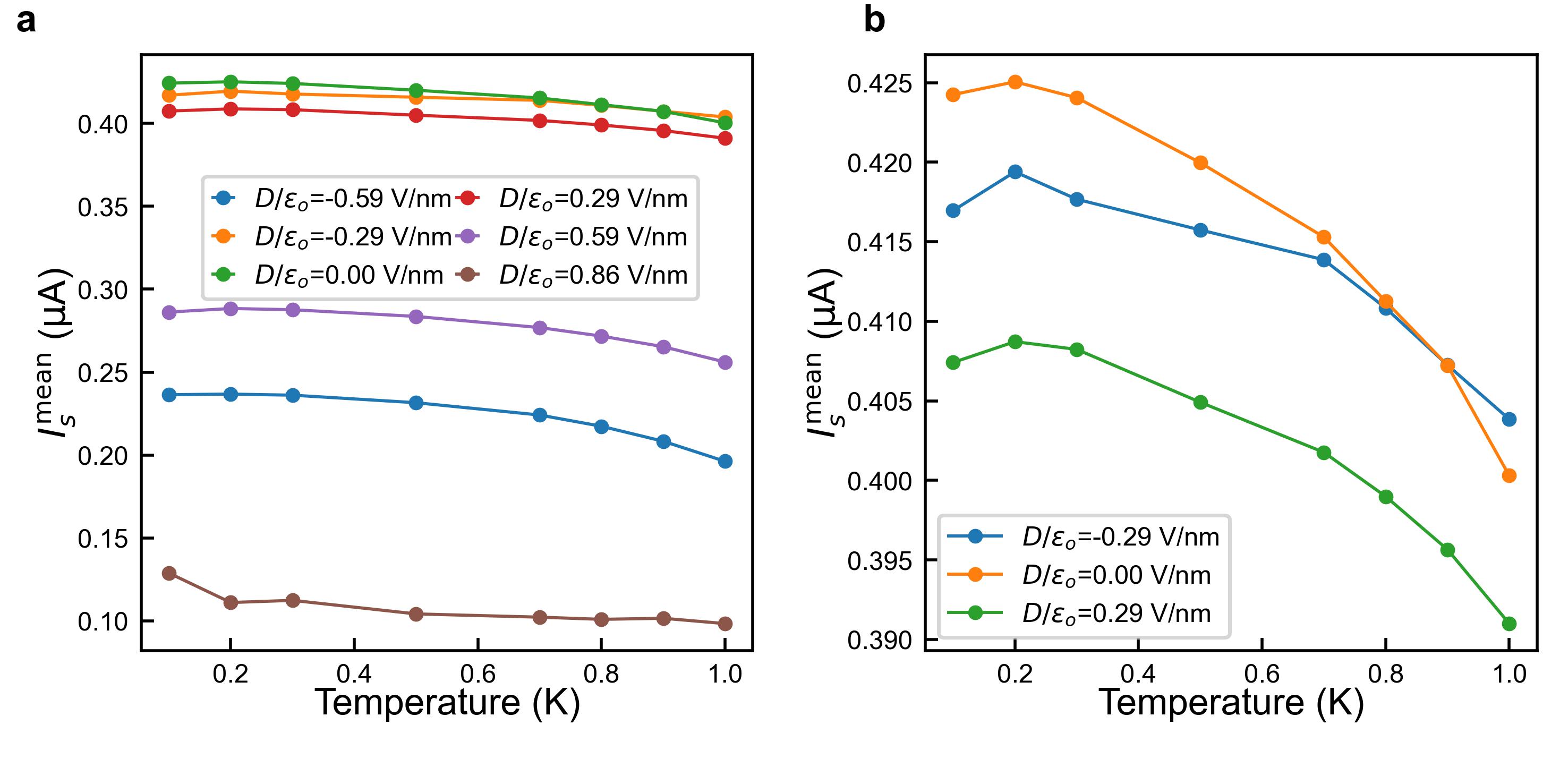}
   \caption{\label{fig:figS14}\textbf{Switching mean current for different perpendicular electric field at $\nu=-2.42$.}
    \textbf{a,} The mean switching current $I_{\mathrm{s}}^{\mathrm{mean}}$ plotted as a function of temperature for different perpendicular electric fields $D/\epsilon_{0}$. 
    \textbf{b,} Zoomed in version of (a) showing the plots for which non-monotonicity in $I_{\mathrm{s}}^{\mathrm{mean}}$ with temperature can be observed.
    }
    \end{figure}

At the optimum filling of $\nu$=-2.42 for the hole-side superconductivity, the $I_{\mathrm{s}}^{mean}$ is studied for switching distributions with increasing temperature. This is then plotted for different perpendicular electric fields $D/\epsilon_{0}$ as shown in Fig.~\ref{fig:figS14}. The perpendicular electric field breaks inversion symmetry in the system and allows for hybridization between different bands leading to change in the bandstructure. The characterisation of the non-monotonic behaviour with different electric fields will allow for possible theories to compare the strengths of the superconducting phase and a competing order as the bandstructure changes with electric field.

\subsection{D. Switching histograms at higher temperature - evolution of bi-modal structure of histograms}

\begin{figure}[h!]
    \centering
    \includegraphics{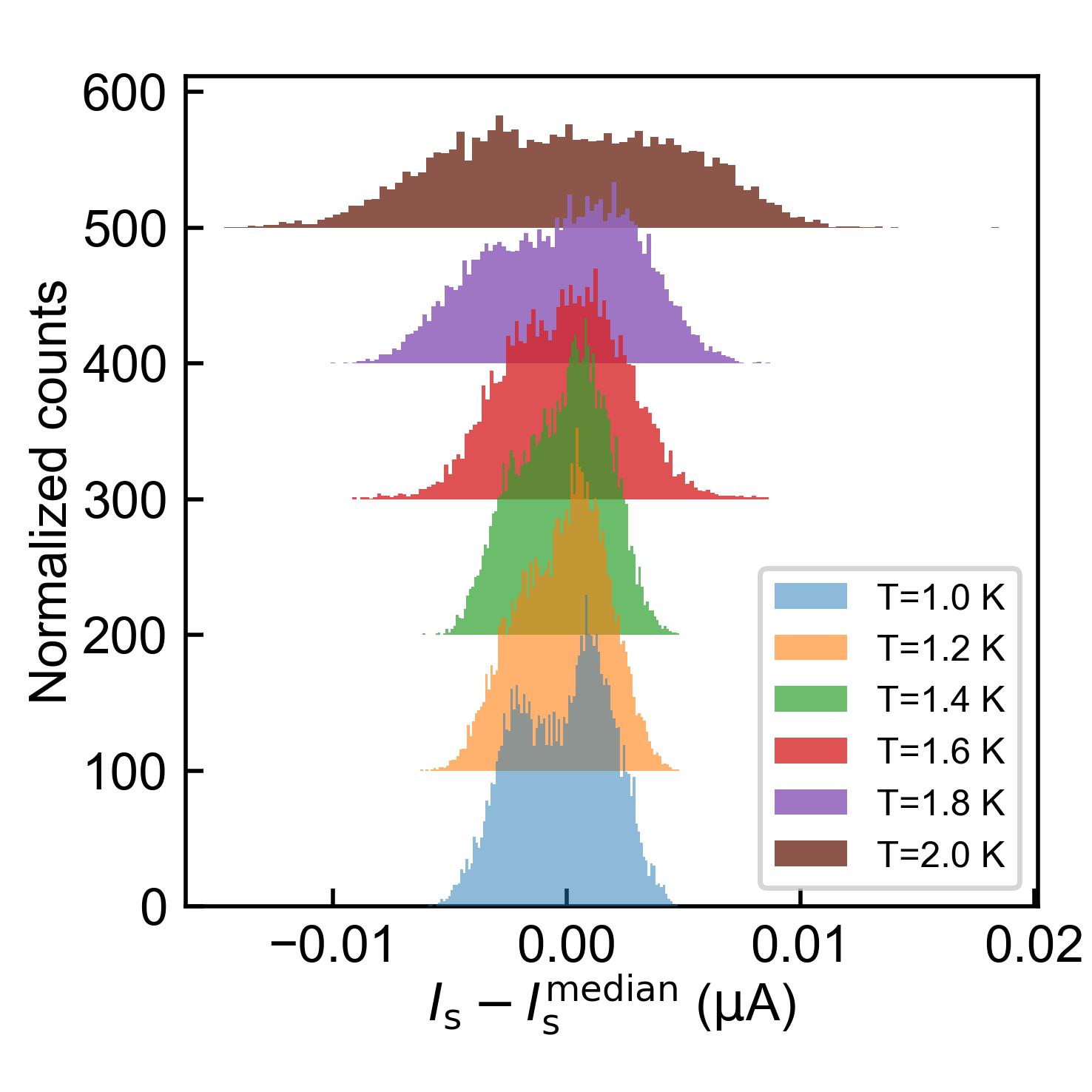}
   \caption{\label{fig:figS15}\textbf{Evolution of the bimodal structure of the histograms with temperature.}
   The switching histograms are plotted for different temperatures at the optimal hole-side filling and electric field ($\nu$, $D/\epsilon_0$) = (-2.42, 0). The x-axis, the median current $I_{\mathrm{s}}^{\mathrm{median}}$ subtracted from the switching current distribution $I_{\mathrm{s}}$, allows us to track the evolution of the bi-modal structure of the distribution into a multi-modal structure with temperature. The histograms for different temperatures are shifted along the y-axis by 100 counts for clarity.    }
\end{figure}

The switching histogram at $\nu$=-2.42, $D/\epsilon_{0}$=0 V/nm and $T$= 1 K develops a substructure of two peaks - bimodal nature. As we discuss in the main manuscript, the bimodal nature of the distribution points to the presence of moir\'e solitons and twistons in the system. Moir\'e solitons and twistons form weak links between the superconducting plaquettes in the system. The weak links forming a network of Josephson junctions (JJs) are penetrated by Josephson vortices. This couples the neighboring JJs. The sub-network of JJs having the vortices will have lower critical currents than the sub-networks not having any vortices. As we increase the temperature, this difference in critical currents becomes more apparent leading to a bi-modal switching distribution. This bi-modal distribution further develops into a very broad distribution as the temperature is increased (as shown in Fig.~\ref{fig:figS15}) and a more complex distribution of vortices is formed in the system. Such bi-modal distributions are also noted in high Tc superconductors due to the coupling of JJs by Josephson vortices and are known as fluxons \cite{krasnov2000comparison}.

\section{XI. CAPACITANCE ESTIMATE WITHIN RCSJ MODEL}\label{sec:sec9}

In the resistively capacitance shunted junction (RCSJ) model, the quality factor is given as
$Q=\omega_{p} R_{\mathrm{N}} C$ where, $\omega_{p}$, is the plasma frequency, $R_{\mathrm{N}}$ is the normal state resistance, and $C$ is the shunt capacitance in the equivalent RCSJ circuit. For our case, $R_{\mathrm{N}}\sim 1~\mathrm{k \Omega}$ and $\omega_{p}=\sqrt{\frac{2 \pi I_{c}}{\Phi_{o}\times C}}$. However, an experimental approximation of the quality factor $Q$ is as $Q_{exp}=\frac{4 I_{c}}{\pi I_{r}}$, where $I_{c}$ is the critical current and $I_{r}$ is the retrapping current. For our MATTG devices, we note no hysteresis in DC I-V curves making $I_{c} \simeq I_{r}$ and $Q\simeq \frac{4}{\pi}$. Equating the $Q_{exp}$ and the RCSJ $Q$, we are able to estimate $C$ to be 1.3~fF. Our measurements on MATTG provide a way to examine the response from the device as an effective JJ.

\section{XII. ESTIMATION OF CROSS-OVER TEMPERATURE FROM MQT TO TA REGIME}

Using again the equation, $\omega_{p}=\sqrt{\frac{2 \pi I_{c}}{\Phi_{o}\times C}}$, where, $\omega_{p}$ is the plasma frequency in the RCSJ model, $I_{c}$ is the critical current of the system, $\Phi_{o}$ is the flux quanta and $C$ is the shunt capacitance within the RCSJ model. In our case, $I_{c}=400~\mathrm{nA}$ and $C =$ 1.3 fF, estimated in section XI. of the SI; making $\omega_{p} =972.029~\mathrm{GHz}$. Using the equation in Likharev \cite{likharev2022dynamics}, we get,
$T_{\mathrm{CO}}=\frac{\hbar \omega_{p}}{2 \pi k_{B}}$, which gives us the cross-over (from macroscopic quantum tunneling regime to thermal activation regime) temperature $T_{\mathrm{CO}}$ to be 1.17 K. This is in the same order of magnitude of the cross-over temperature observed in our experiment of $\simeq1$ K, as shown in Fig. 2f of the main manuscript.
\begin{figure}[h!]
    \centering
    \includegraphics{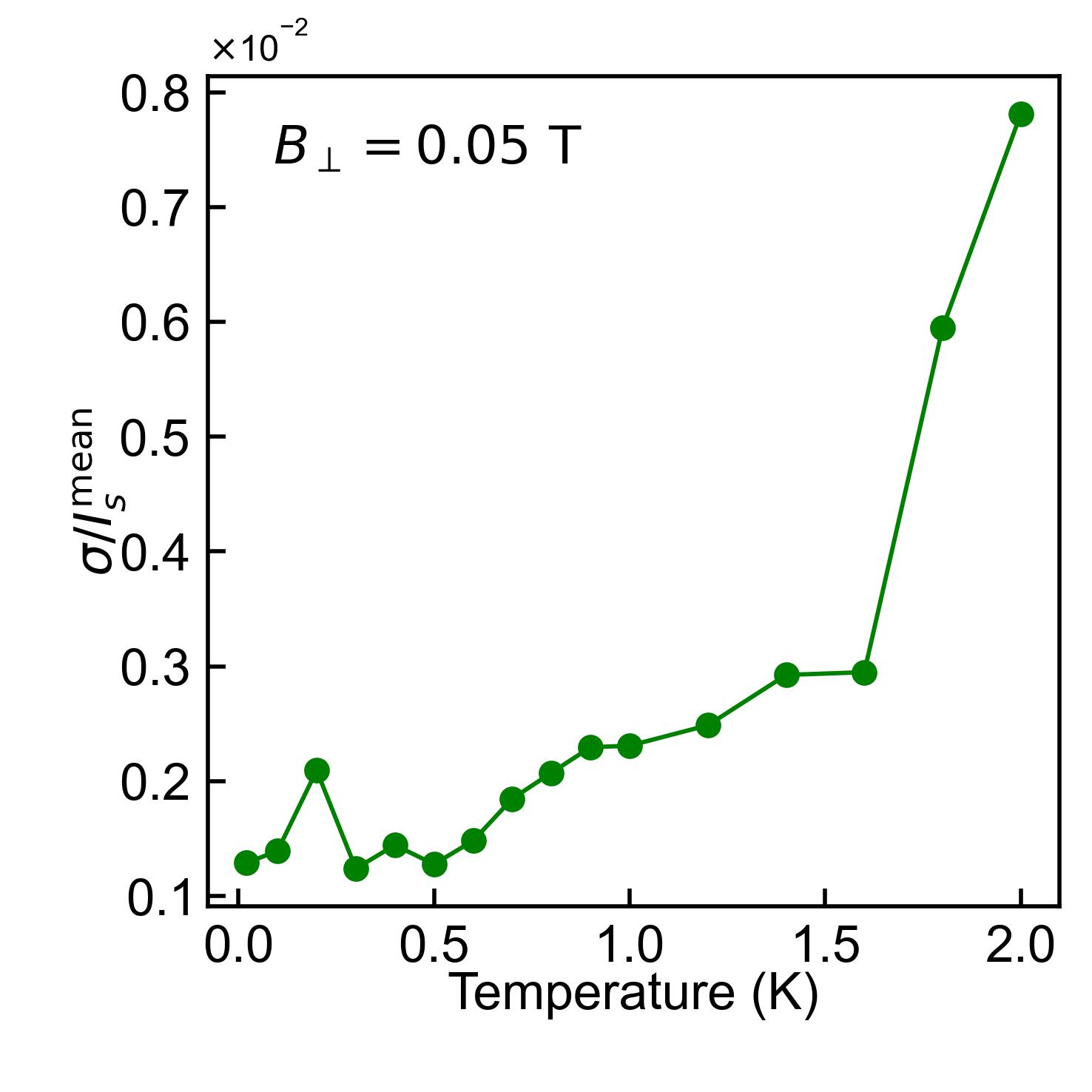} 
   \caption{\label{fig:figS16}\textbf{Suppression of cross-over temperature $T_{\mathrm{CO}}$ in the presence of finite perpendicular magnetic field.}
    The standard deviation divided by the mean switching current $\sigma/I_{\mathrm{s}}^{\mathrm{mean}}$ plotted as a function of temperature. The cross-over temperature $T_{\mathrm{CO}}$ marking the transition from the MQT to the TA regime in a network of JJs is suppressed to $T_{\mathrm{CO}}\simeq$ 0.5 K due to the application of a finite perpendicular magnetic field $B_{\perp}$= 0.05 T. In contrast, at $B_{\perp}$= 0 T, the $T_{\mathrm{CO}}\simeq$1 K as can be noted in Fig. 2f  of the main manuscript.}
    \end{figure}
Additional evidence of the JJ network picture in MATTG is provided by the suppression of the cross-over temperature $T_{\mathrm{CO}}$ on the application of a small and finite perpendicular magnetic field \cite{krasnov2005collapse} (see Fig.~\ref{fig:figS16} where the $T_{\mathrm{CO}}$ has suppressed to a value of $\simeq0.5$ K).

\newpage

\section{XIII. THERMAL CYCLING IN THE PRESENCE AND ABSENCE OF IN-PLANE MAGNETIC FIELD}

\begin{figure}[h!]
    \centering
    \includegraphics{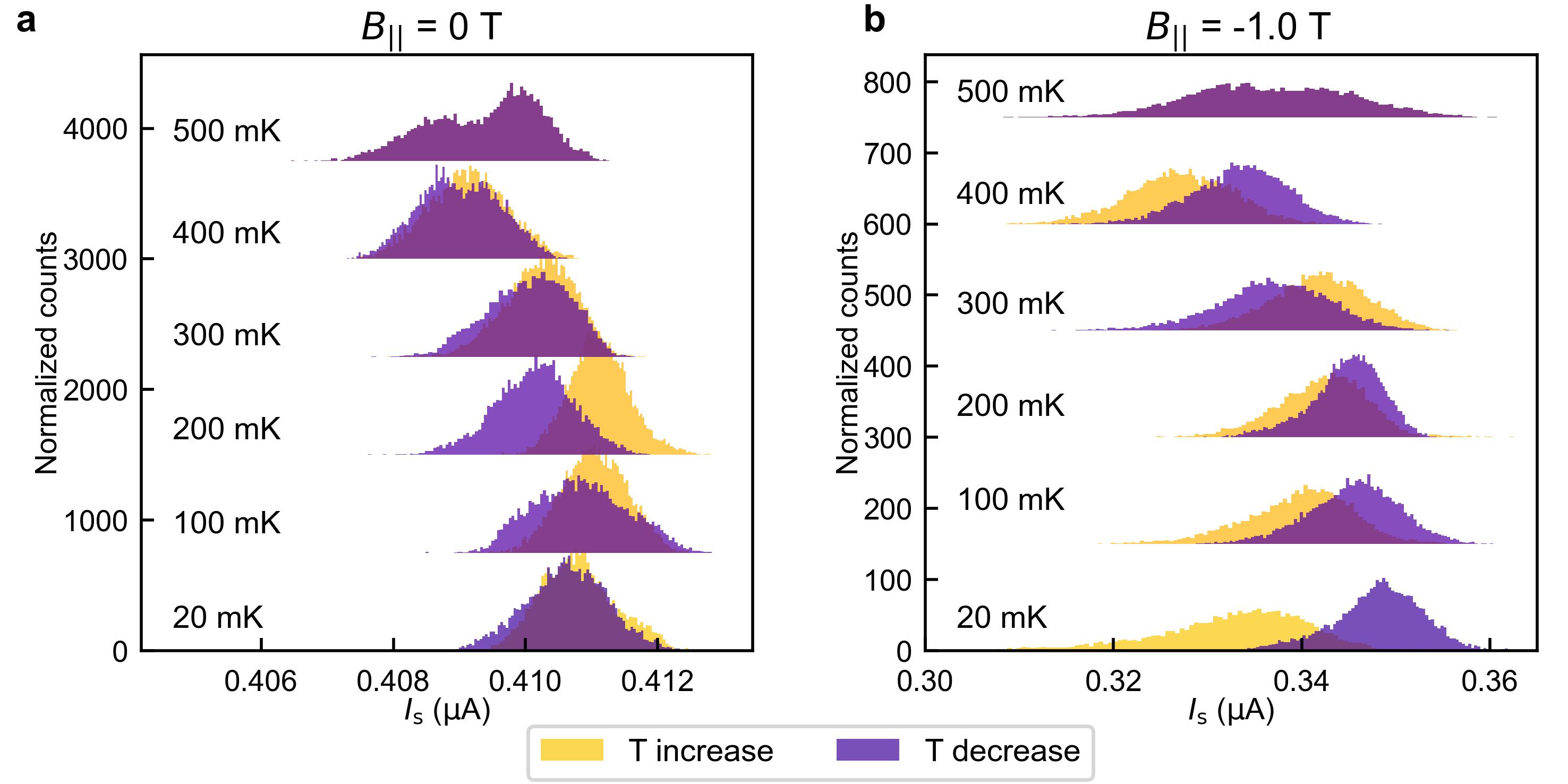}
   \caption{\label{fig:figS17}\textbf{Thermal cycling in the presence and absence of magnetic field.} 
   \textbf{a,b,}  The switching histograms at (a) $B_{\mathrm{||}}$=0 T and (b) $B_{\mathrm{||}}$=-1 T show history dependence for thermal cycling up to 500 mK. Each histogram pair, yellow for the increasing leg and purple for the decreasing leg of the cycle are at a particular temperature. Each pair is shifted along the y-axis by 750 units in (a) and 150 units in (b) for clarity.
   }
   \end{figure}

Fig.~\ref{fig:figS17} shows the switching distributions at the optimum hole-side doping. The switching distributions are plotted for different temperatures during thermal cycling from 20 mK to 500 mK and back. In the absence of in-plane magnetic field, the switching distributions for a particular temperature in the heating and cooling cycle lie almost on top of each other, except for 200 mK distribution, which is also the temperature at which we note the non-monotonicity of $I_{\mathrm{s}}^{\mathrm{mean}}$.
On the contrary, in the presence of in-plane magnetic field $B_{\mathrm{||}}$=-1 T, there is an appreciable change in switching distributions between the heating and the cooling cycle at each temperature. The most remarkable are the histograms at 20 mK. In \ref{fig:figS17}a, the histograms at 20 mK overlap while in \ref{fig:figS17}b, the 20 mK histograms not only shift in the $I_{\mathrm{s}}$ axis but also differ in shape. This points towards the complex energetics in the system due to the interplay of the competing orders - magnetic and superconducting phases.

\section{XIV. HYSTERESIS WITH PERPENDICULAR FIELD}

\begin{figure}[h!]
    \centering
    \includegraphics{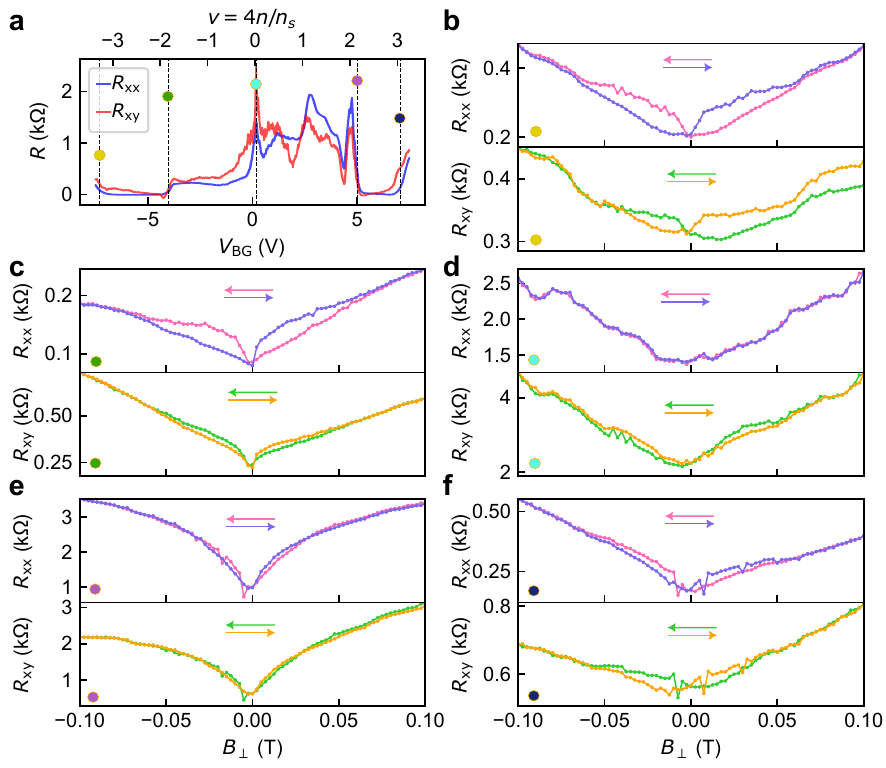}
    \caption{\footnotesize \label{fig:figS18} {\textbf{Hysteresis with perpendicular field}
    \textbf{a,}~Longitudinal resistance $R_{\mathrm{xx}}$ and Hall resistance $R_{xy}$ as a function of the bottom gate voltage $V_{\mathrm{BG}}$ at $B$=0. The top axis represents the corresponding filling factor $\nu$. The colored circles identify particular fillings for the next measurements.
    \textbf{b-f}~$R_{\mathrm{xx}}$ and $R_{\mathrm{xy}}$ plotted as a function of perpendicular magnetic field $B_{\mathrm{\perp}}$ as it is swept in the two directions indicated by the arrows. The symbol on the bottom left identifies the filling factor from \textbf{a}. The magnetic field is swept at a rate of $\sim$0.006 T/min.
    }}
\end{figure}

In Fig.~\ref{fig:figS18}, we show the hysteresis in Hall resistance $R_{xy}$, measured simultaneously with longitudinal resistance $R_{xx}$ in a perpendicular magnetic field $B_{\mathrm{\perp}}$.
Fig.~\ref{fig:figS18}a shows the fillings corresponding to the hysteresis data in Fig.~\ref{fig:figS18}b-f. 
A hysteresis in Hall resistance $R_{xy}$ shows further evidence of broken time-reversal symmetry due to magnetic order, in close proximity to the superconducting state.

\section{XV. SUPERCONDUCTING DIODE EFFECT}

\begin{figure}[h!]
    \centering
    \includegraphics{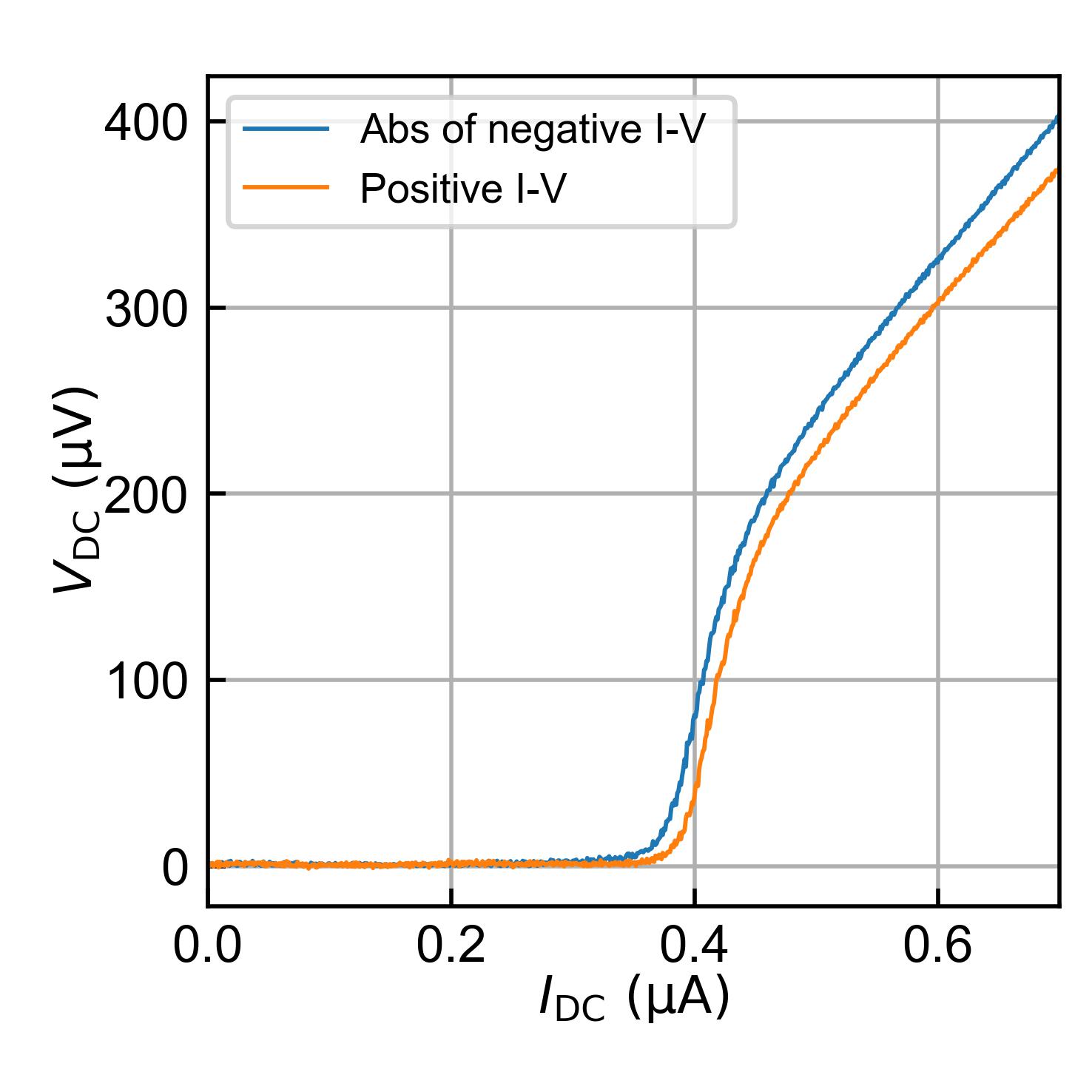}
   \caption{\label{fig:figS19}\textbf{Superconducting diode effect.}
    DC I-V line slice at $\nu=-2.42$, $D/\epsilon_0=0$ V/nm comparing the I-V in the positive current bias and the absolute value of negative current bias, showing the presence of small $\simeq1.2\%$ diode effect.
    }
    \end{figure}

The diode effect can be characterized by the asymmetry, which is the normalized difference in critical currents of negative and positive current bias, $\Delta I_{c}=\frac{I_{c}^{+}-|I_{c}^{-}|}{I_{c}^{+}+|I_{c}^{-}|}$. Extracting the values from Fig.~\ref{fig:figS19}, we get a diode effect asymmetry of 1.2\%.

\section{XVI. SUPERFLUID STIFFNESS ANALYSIS}

\begin{figure}
    \centering
    \includegraphics{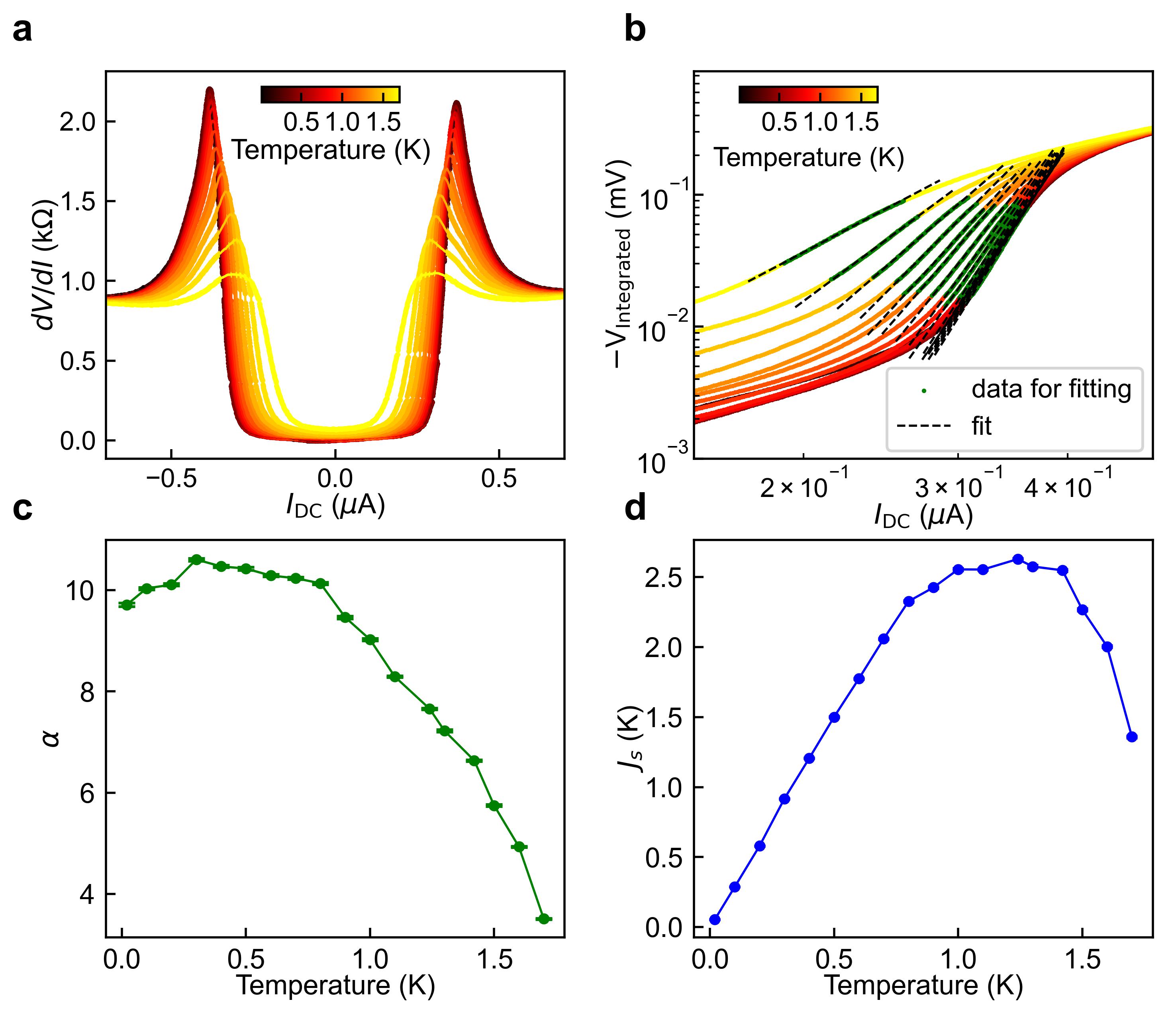}
    \caption{\label{fig:figS20}\textbf{Superfluid stiffness analysis at higher currents corresponding to depairing of Cooper pairs.}
    \textbf{a,}~Differential resistance $dV/dI$ vs $I_{\mathrm{DC}}$ curves at temperatures from 20~mK to 1.7~K.
    \textbf{b,}~ Integrated DC I-V curves from differential resistance $dV/dI$ vs $I_\mathrm{{DC}}$ curves, fitted in log-log plot for their non-linear exponent $\alpha$ at currents near the critical current.
    \textbf{c,}~Extracted exponent $\alpha$ from the analysis at near critical current values as a function of temperature $T$. $\alpha$ crosses the BKT transition point of 3, and gives us a false impression of a BKT transition.
    \textbf{d,}~The BKT relations allow us to infer the values of superfluid stiffness $J_{\mathrm{s}}$ from the high-current analysis which also gives us incorrect values. This is so as the exponent captured in the high-current analysis is not related to the underlying BKT physics but to inhomogeneities in the system and the depairing of Cooper pairs.
    }

\end{figure}

The analysis of superfluid stiffness depends on the extraction of a non-linear exponent $\alpha$ from the DC I-V curves. The DC I-V curves used for analysis are obtained by numerical integration of the differential resistance $dV/dI$ vs $I_\mathrm{{DC}}$ curves. This is done because the lock-in amplifier used for the measurement of the differential resistance has better sensitivity. It gives us smoother curves than the raw DC I-V data and is suitable for extraction of $\alpha$. The DC I-V curves are plotted in a log-log scale and a straight line is fit to the data at a low current value, one order less than the critical current. Calculation suggests typical currents as $I^{*}\mathrm{[A]}\simeq 2.76 \times 10^{-8}T_{\mathrm{BKT}}\mathrm{[K]}$ \cite{benfatto2009broadening,baity2016effective}, where $T_{\mathrm{BKT}}$ is the Berezinskii – Kosterlitz – Thouless (BKT) transition temperature. The slope extracted out of this straight-line fit is the said non-linear exponent $\alpha$.

The analysis is also repeated for values of higher current, near the critical current values for comparison. The results of the analysis are presented in Fig.~\ref{fig:figS20}. This analysis captures the depairing of Cooper pairs and not the vortex-antivortex motion and thus the exponent extracted, although crosses the BKT transition mark of 3, gives us a false impression of a BKT transition. This further justifies the importance of carrying out this analysis at a lower current, such that it captures the effect of depairing of the vortex-antivortex pairs only.

\end{document}